\documentclass[12pt,reqno]{amsart}
\usepackage{latexsym, amsfonts, amsmath, amsthm, amssymb, amscd, color}
\usepackage{pstricks}
\usepackage{pst-optic}
\usepackage{caption}
\usepackage{listings}
\usepackage{graphics}    
\usepackage{graphicx}
\usepackage{color}
\usepackage{mathrsfs}
\usepackage{xparse}
\usepackage{enumerate}
\usepackage{verbatim}
\usepackage{gensymb}
\usepackage{float}

\DeclareFontFamily{U}{mathx}{\hyphenchar\font45}
\DeclareFontShape{U}{mathx}{m}{n}{
      <5> <6> <7> <8> <9> <10>
      <10.95> <12> <14.4> <17.28> <20.74> <24.88>
      mathx10
      }{}
\DeclareSymbolFont{mathx}{U}{mathx}{m}{n}
\DeclareFontSubstitution{U}{mathx}{m}{n}
\DeclareMathAccent{\widecheck}{0}{mathx}{"71}
\DeclareMathAccent{\wideparen}{0}{mathx}{"75}

\newtheorem*{thm*}{Theorem}
\newtheorem{thm}{Theorem}
\newtheorem{prop}[thm]{Proposition}

\newtheorem{cor}[thm]{Corollary}

\theoremstyle{remark}
\newtheorem{rmk}{Remark}

\numberwithin{equation}{section}

\newcommand{\Hreg}[4]{\ensuremath{H_{#1,#2}^{#4,#3}}}
\newcommand{\Vreg}[3]{\ensuremath{V_{#1,#2}^{#3}} }

\newcommand{\HPreg}[4]{\ensuremath{\widehat{H}_{#1,#2}^{#4,#3}}}

\newcommand{\HMreg}[4]{\ensuremath{\widecheck{H}_{#1,#2}^{#4,#3}}}

\newmuskip\pFqskip
\mathchardef\pFcomma=\mathcode`, 

\newcommand*\pFq[5]{%
  \begingroup
  \begingroup\lccode`~=`,
    \lowercase{\endgroup\def~}{\pFcomma\mkern\pFqskip}%
  \mathcode`,=\string"8000
  {}_{#1}F_{#2}\biggl[\genfrac..{0pt}{}{#3}{#4};#5\biggr]%
  \endgroup
}

\begin{document}
\title{Interactions between interleaving holes in a sea of unit rhombi}
\author[T. Gilmore]{Tomack Gilmore}
  \begin{abstract}
Consider a family of collinear, equilateral triangular holes of any even side 
length lying within a sea of unit rhombi. The results presented below show that 
as the distance between the holes grows large, the interaction between them may 
be approximated, up to a multiplicative constant, by taking the exponential of 
the negative of the electrostatic energy of the system obtained by viewing the 
holes as a set of point charges, each with a signed magnitude given by a 
certain statistic. Furthermore it is shown that the interaction between a 
family of left pointing collinear triangular holes and a free boundary may be 
approximated (again up to some multiplicative constant) by taking the 
exponential of the negative of the electrostatic energy of the system obtained 
by considering the holes as a set of point charges and the boundary a straight 
equipotential conductor. These two differing systems of point charges can be 
related via the method of image charges, a well-known physical law that also 
surfaces in the following mathematical analysis of enumeration formulas that 
count tilings of certain regions of the plane by unit rhombi. 
 \end{abstract}
 \maketitle
\section{Introduction}\label{sec:Intro}
Interactions between holes (or gaps) in two dimensional dimer systems were 
first considered by Fisher and Stephenson, whose seminal paper~\cite{FishSteph} 
examined three types of interaction: the interaction between two 
dimers; the interaction between two monomers; and the interaction between a 
dimer and a fixed boundary (that is, an edge or a corner). While this work 
focused exclusively on interactions between holes in the square lattice, 
Kenyon~\cite{Keny1} later generalised the first of these 
interaction types to an arbitrary number of dimer gaps on both the square and 
hexagonal lattices. Kenyon, Okounkov, and Sheffield~\cite{KeOkSh} then extended 
these results even further to include general bipartite planar lattices.

Interactions between non-dimer gaps on the hexagonal lattice (in particular 
gaps consisting of a pair of monomers) have been studied 
extensively by Ciucu~\cite{CiuRot}\cite{CiuRand}\cite{CiuScale}\cite{CiuEmer}, 
establishing therewith close (conjectural) analogies between such 
interactions and two dimensional electrostatic phenomena. More specifically, 
Ciucu conjectures that the asymptotic interaction between non-dimer holes in a 
two 
dimensional dimer system on the hexagonal lattice is, up to a multiplicative 
constant, inversely 
proportional to the product of the pairwise distances between the holes raised 
to some exponential power that is determined by each pair of holes. Such an 
interaction 
is said to be governed by Coulomb's law for electrostatics since it may be 
obtained by taking the exponential of the negative of the electrostatic energy 
of the two-dimensional system of physical charges obtained by considering the 
holes as point charges of a certain magnitude and sign. A more formal statement 
of this conjecture together with a discussion of the evidence in support of it 
may be found in Ciucu's excellent survey paper~\cite{CiuPNAS}. 
Although it has been shown to hold for a very general class of holes in 
dimer systems that are embedded on tori~\cite{CiuScale} and also for a 
certain class of holes in planar dimer systems~\cite{CiuRand}, a 
complete proof of the conjecture remains elusive.

Further to the above mentioned interactions, a wholly (or perhaps, 
\emph{hole-ly}) new type of non-dimer interaction was presented more recently 
in~\cite{CiuKrattInt}: that of the interaction between a triangular hole and a 
so-called ``free" boundary. Such an interaction also parallels certain 
physical phenomena, namely it appears to behave in analogy to the 
attraction of an electric charge to a straight line conductor. The 
author of the present work showed in~\cite{TG1} that similar behaviour may also 
be observed for a triangular hole that has been rotated 180$\degree$ and thus 
points toward the boundary. Somewhat mysteriously, it seems the 
orientation of the hole has a direct effect on the interaction between the hole 
and the free boundary (a physical interpretation of this discrepancy has yet 
to be realised).

\begin{figure}\centering
\includegraphics[scale=0.75]{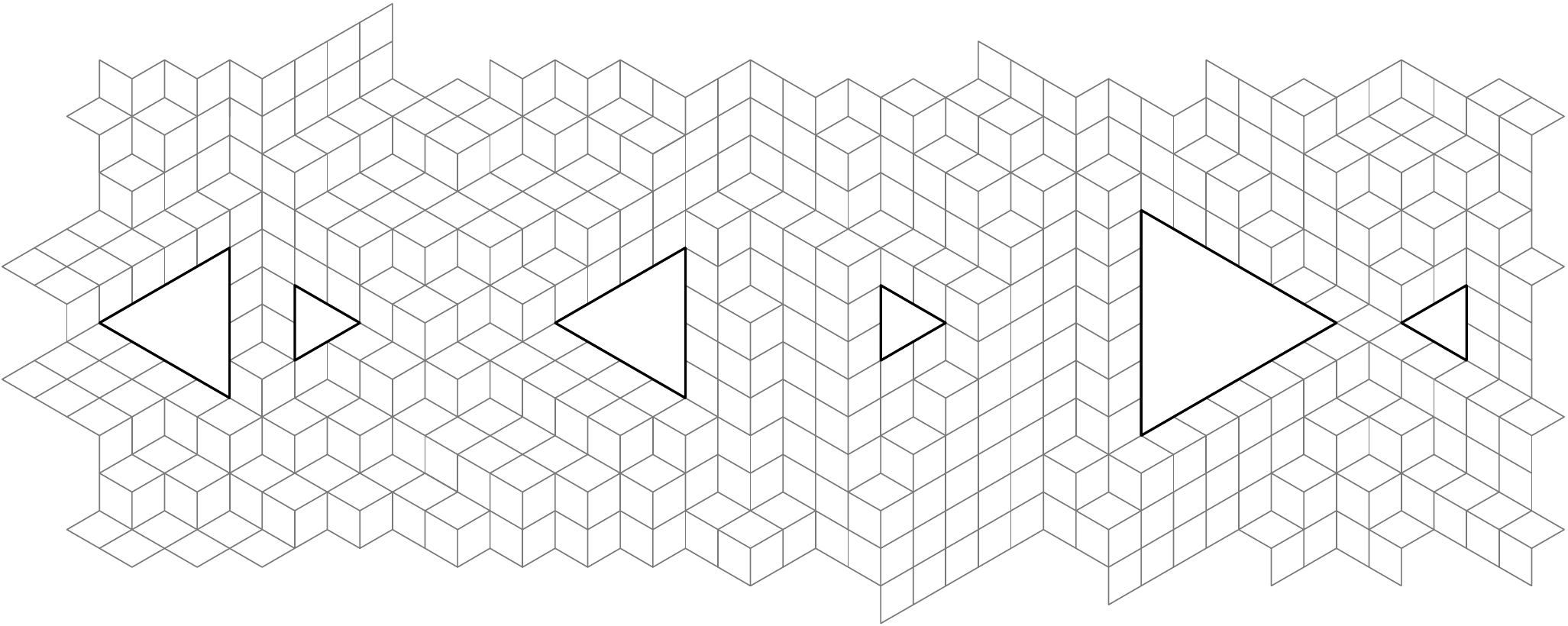}\caption{Part of a rhombus tiling of 
the 
plane, containing a set of horizontally collinear triangular holes of even side 
lengths where the sum 
of the charges of each hole is zero.}\label{fig:CorrFull}
\end{figure}

The contribution of the current article is twofold. Firstly, the 
interaction of an 
entirely new class of triangular holes is established, namely the interaction 
between \emph{interleaving} holes of any even side length that lie along a 
horizontal line within a sea 
of unit rhombi (here 
interleaving means that all holes that 
point in one direction do not necessarily all lie to one side of all holes that 
point in the opposite direction, see Figure~\ref{fig:CorrFull}). It would 
appear that aside from the aforementioned results obtained by embedding tilings 
on tori~\cite{CiuScale}, such interactions for the planar case have yet to be 
treated in the literature. Theorem~\ref{thm:MainThm} below shows that under 
certain conditions the planar case does indeed agree with Ciucu's tori result, 
however the results stated in Section~\ref{sec:Asymp} show that this 
type of interaction depends somewhat delicately on the rate at which the 
boundaries of the plane approach infinity.

Secondly, the interaction between a set of 
left pointing triangular holes and a free boundary is established, thus 
generalising the earlier work of Ciucu and 
Krattenthaler~\cite{CiuKrattInt} to include any (finite) number of left 
pointing 
triangular holes of any even side length. Once again it is shown 
that such interactions are in some sense governed by Coulomb's law, since they 
may be approximated by taking the exponential of the negative 
of the electrostatic energy of the system obtained by viewing the 
holes as a set of point charges and the free boundary a straight equipotential 
conductor. According to the method of images~\cite[Chapter 6]{Feyn} the 
electrostatic energy of such a system is half the electrostatic energy of the 
system obtained by replacing the conductor with imaginary charges 
(these charges of opposite signed magnitude are obtained by reflecting the 
original charges through the 
conductor). One sees this well-known physical law 
surfacing through the purely mathematical analysis of certain enumeration 
formulas that count tilings of certain regions of the plane by unit rhombi, 
thus adding further support to the on-going electrostatic program of Ciucu.

\section{Set-up and Results}
The main results presented in this article concern \emph{interactions} between 
holes in two dimensional \emph{dimer systems}. In the spirit of Fisher and 
Stephenson~\cite{FishSteph}, suppose $\mathcal{R}_n$ is a subgraph (with size 
parametrised by $n$) of some two dimensional lattice with a fixed set of 
vertices (indexed by the set 
$\mathcal{H}$) removed from its interior. Denote this region containing a set 
of holes 
$\mathcal{R}_n\setminus\mathcal{H}$. A 
perfect matching between vertices in $\mathcal{R}_n\setminus\mathcal{H}$ is 
also 
known as a \emph{dimer covering} of $\mathcal{R}_n\setminus\mathcal{H}$, and 
the 
set of all coverings of this region is known as a \emph{dimer system}. As $n$ 
tends to infinity, dimer coverings of $\mathcal{R}_n\setminus\mathcal{H}$ 
become dimer coverings of the entire plane (in other words, sending $n$ to 
infinity yields a set of holes that lie within a sea of dimers, see 
Figure~\ref{fig:CorrFull}), and the 
\emph{interaction} between the fixed holes (otherwise known as the 
\emph{correlation function} of the holes) in the dimer system is defined to be
\begin{equation}\label{eq:CorrDef}\omega_{\mathcal{R}}(\mathcal{H})=\lim_{
n\to\infty } \frac { M(\mathcal { R }
_n\setminus\mathcal{H})}{M(\mathcal{R}_n)},\end{equation}
where $M(\mathcal{R})$ denotes the number of dimer coverings (equivalently, 
perfect matchings) of the region $\mathcal{R}$.
\begin{figure}[t]\centering
\includegraphics[scale=0.3]{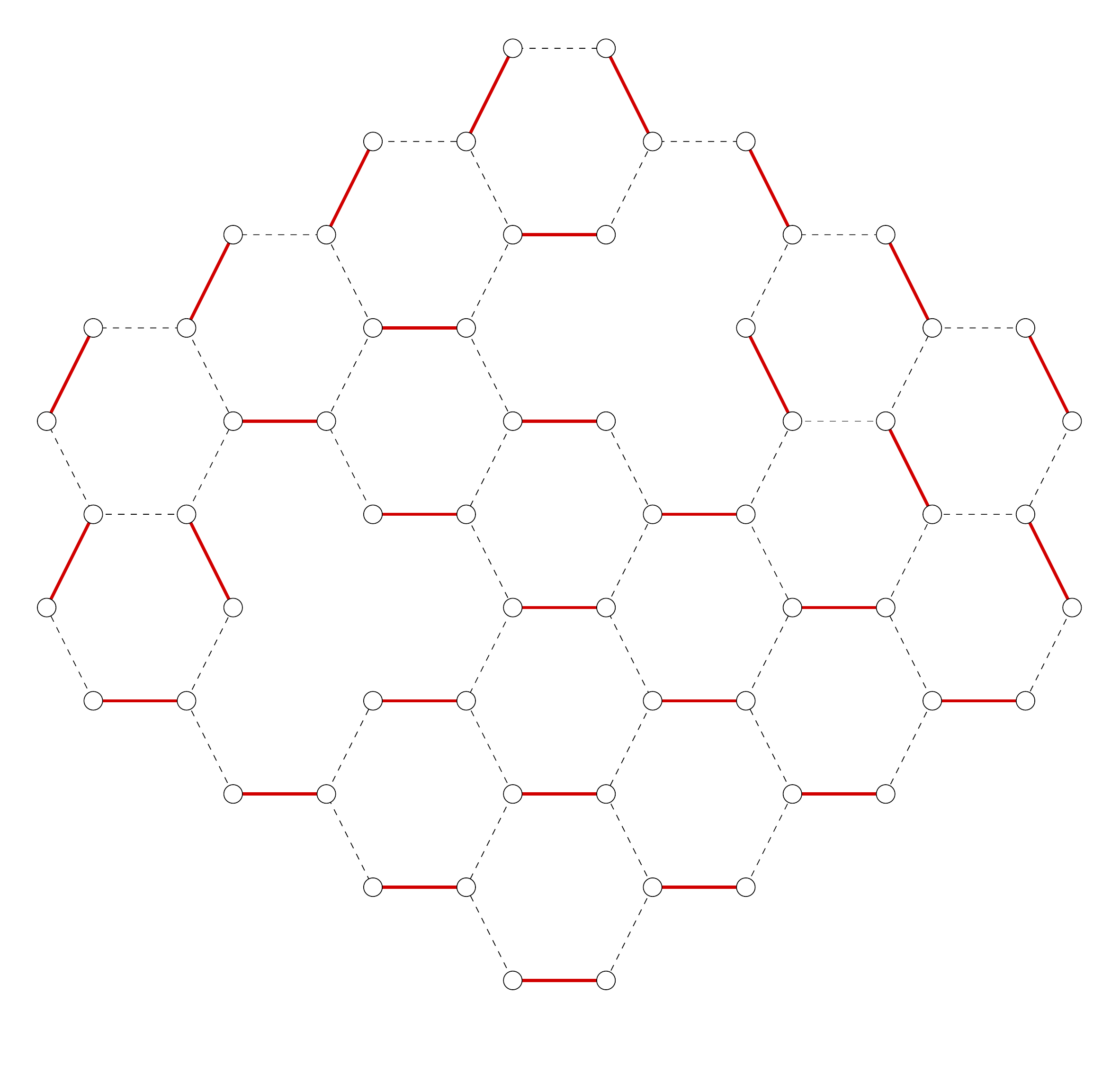}
\includegraphics[scale=0.3]{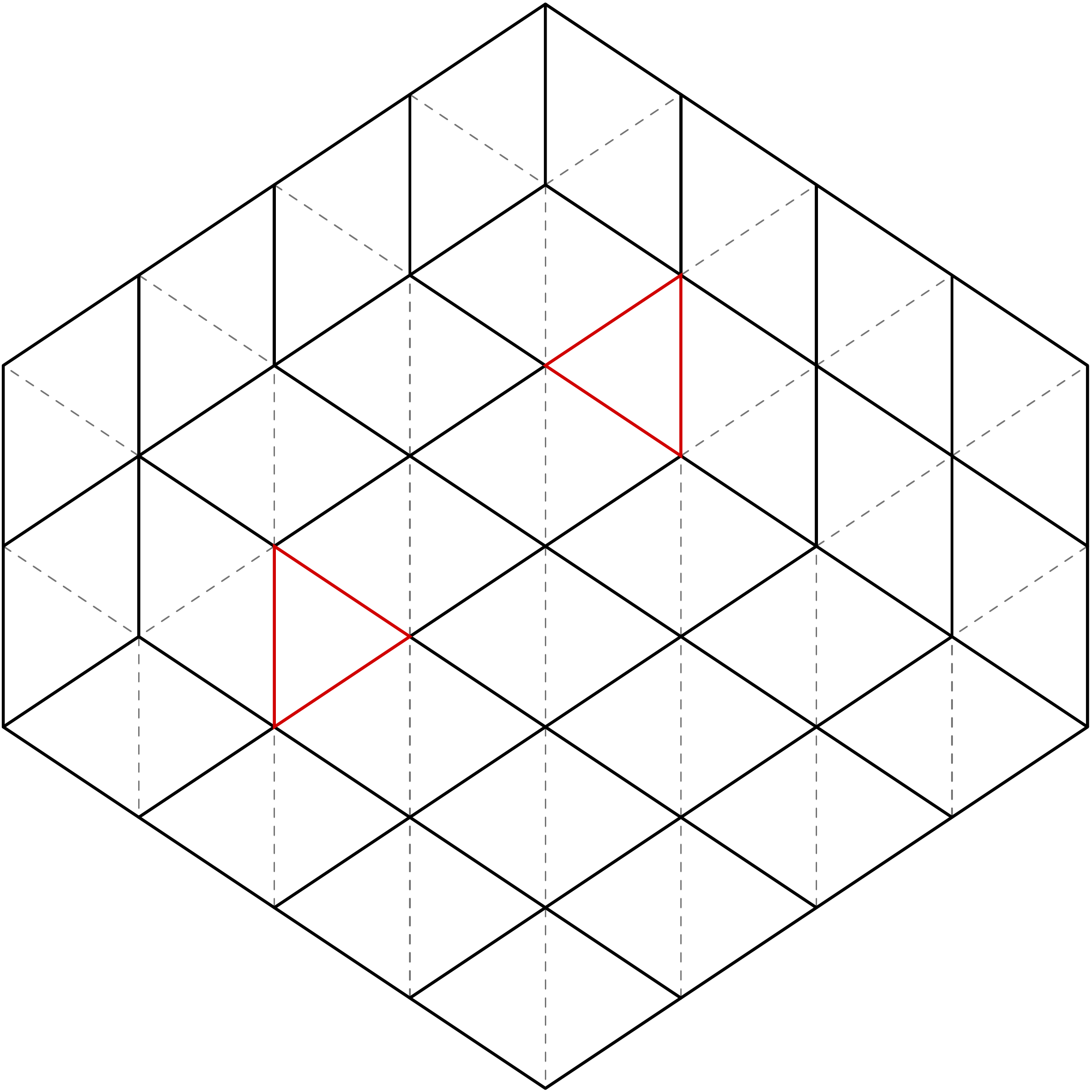}
\caption{A dimer covering of a holey subregion of the hexagonal lattice, left, 
and the corresponding rhombus tiling on the triangular lattice, 
right. The unit triangles in red on the right correspond to the vertices 
that have been removed from the interior of the subgraph on the 
left.}\label{fig:dimer}
\end{figure}
This paper focuses on dimer systems on the planar hexagonal 
lattice, 
$\mathscr{H}$, considered in terms of its ``dual", that is, the 
planar triangular lattice consisting of unit triangles, $\mathscr{T}$, drawn so 
that one of the families of lattice lines is vertical. In this context a 
matching between two neighbouring vertices of $\mathscr{H}$ corresponds to 
joining a pair of unit triangles in $\mathscr{T}$ that share precisely one 
edge, 
therefore a dimer covering of $\mathscr{H}$ (from which a finite number of 
vertices may have been removed) corresponds to a tiling of the plane by unit 
rhombi (where a corresponding set of unit triangles have been removed). 
An example of a dimer covering of a subregion of $\mathscr{H}$ and its 
corresponding rhombus tiling on $\mathscr{T}$ may be found 
in Figure~\ref{fig:dimer}.

It was conjectured by Ciucu~\cite{CiuPNAS} in 2008 that the interaction between 
any set of holes, $\mathcal{H}$, that lie far apart within a sea of unit rhombi 
is 
asymptotically equal to
$$\prod_{h\in\mathcal{H}}\tilde{\omega}(h)\prod_{1\le 
i<j\le|\mathcal{H}|}d(h_i,h_j)^{\frac{1}{2}q(h_i)q(h_j)},$$
where $\tilde\omega(h)$ is a constant dependent on each individual hole, $q(h)$ 
is the \emph{charge}\footnote{The charge of $h$ is a statistic on the hole 
given by the number of right pointing unit triangles that comprise it 
minus the number of left pointing ones. For example a right pointing 
triangular hole of side length two has charge $2$, whereas a left pointing hole 
of the same size has charge $-2$.} of the hole 
$h$, and $d(h_i,h_j)$ is the Euclidean distance between the holes $h_i$ and 
$h_j$ indexed by $\mathcal{H}$. Although this conjecture remains open, the main 
result of this paper 
shows that it holds for a general family of holes that appear to have not yet 
been considered in any of the literature, namely triangular holes of any even 
side length that are \emph{horizontally collinear} (that is, they lie on 
a horizontal line about which they are symmetrically distributed), where the 
sum of the charges of the holes is zero.\footnote{It shall be assumed that all 
sets of holes considered in this paper satisfy this condition.} An example of 
such a family of holes may be found in Figure~\ref{fig:CorrFull}.
\begin{figure}[t]\begin{center}
\includegraphics[scale=0.40]{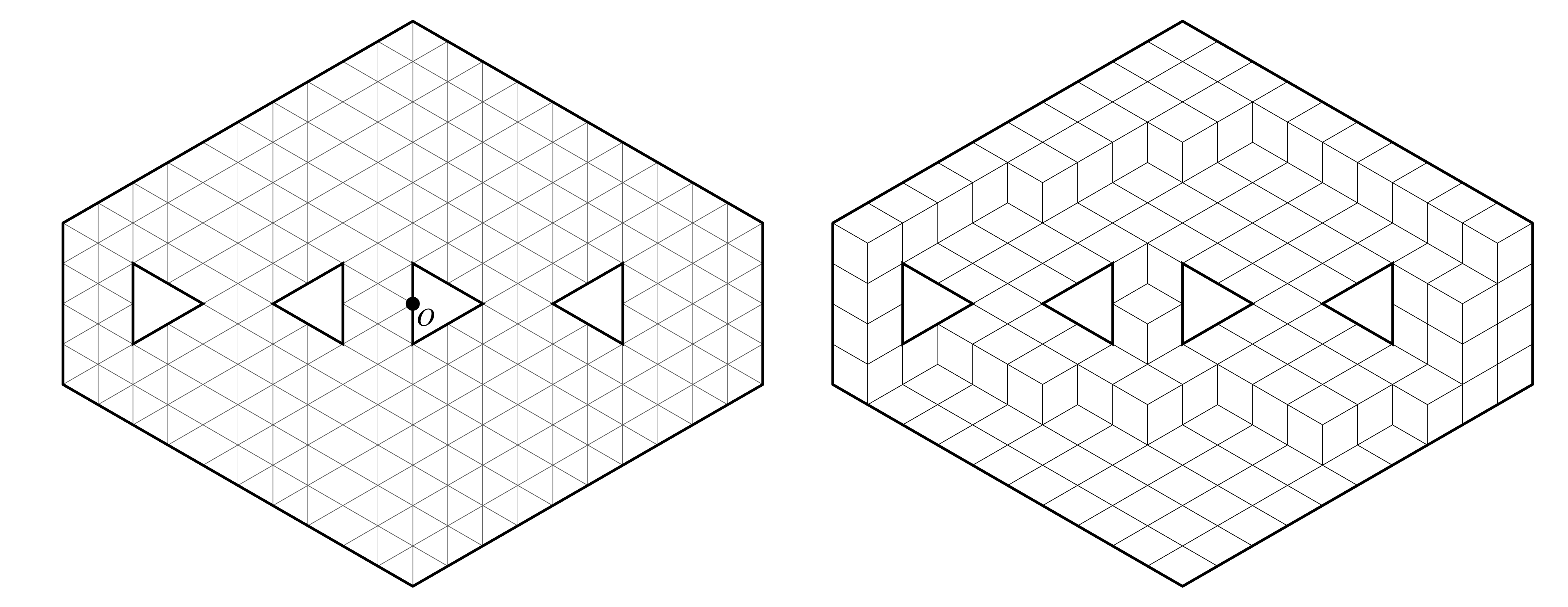}\caption{The holey 
hexagon \Hreg{10}{4}{\{0,-8\}}{\{-2,6\}}, left, and a 
rhombus tiling of the same region, right.}\label{fig:FullTile}\end{center}
\end{figure}
\begin{thm}\label{thm:MainThm}
The interaction between a set $\mathcal{H}$ of horizontally collinear 
triangular 
holes of any even side length is asymptotically
$$\omega_H(\mathcal{H})\thicksim\prod_{h\in\mathcal{H}}C_{h}\prod_{
1\le j< i\le |\mathcal{H}|}d(h_i,h_j)^{\frac{1}{2}q(h_i)q(h_j)}$$
as the distance between the holes $h_i$ and $h_j$ in $\mathcal{H}$ becomes 
large, where 
$$C_h=\prod_{s=0}^{\frac
{ 1 } { 2 } |q(h)|-1 } \frac{3^{s+1/2}}{2\pi}\Gamma(s+1)^2.$$
\end{thm}
\begin{rmk}\label{rmk:Coulomb1}
The above theorem shows that the  
interaction between sets of interleaving, collinear triangular holes of any 
even side length may be approximated, up to a multiplicative constant, by 
taking the exponential of the negative of the electrostatic energy of the system 
obtained by viewing each hole as a point charge with signed magnitude 
given by the statistic $q$. A 
more detailed discussion of the close analogies between rhombus tilings of 
regions containing holes and 
certain electrostatic phenomena may be found in~\cite{CiuPNAS} 
and~\cite{CiuKrattInt}.
\end{rmk}
\begin{figure}
\includegraphics[scale=0.55]{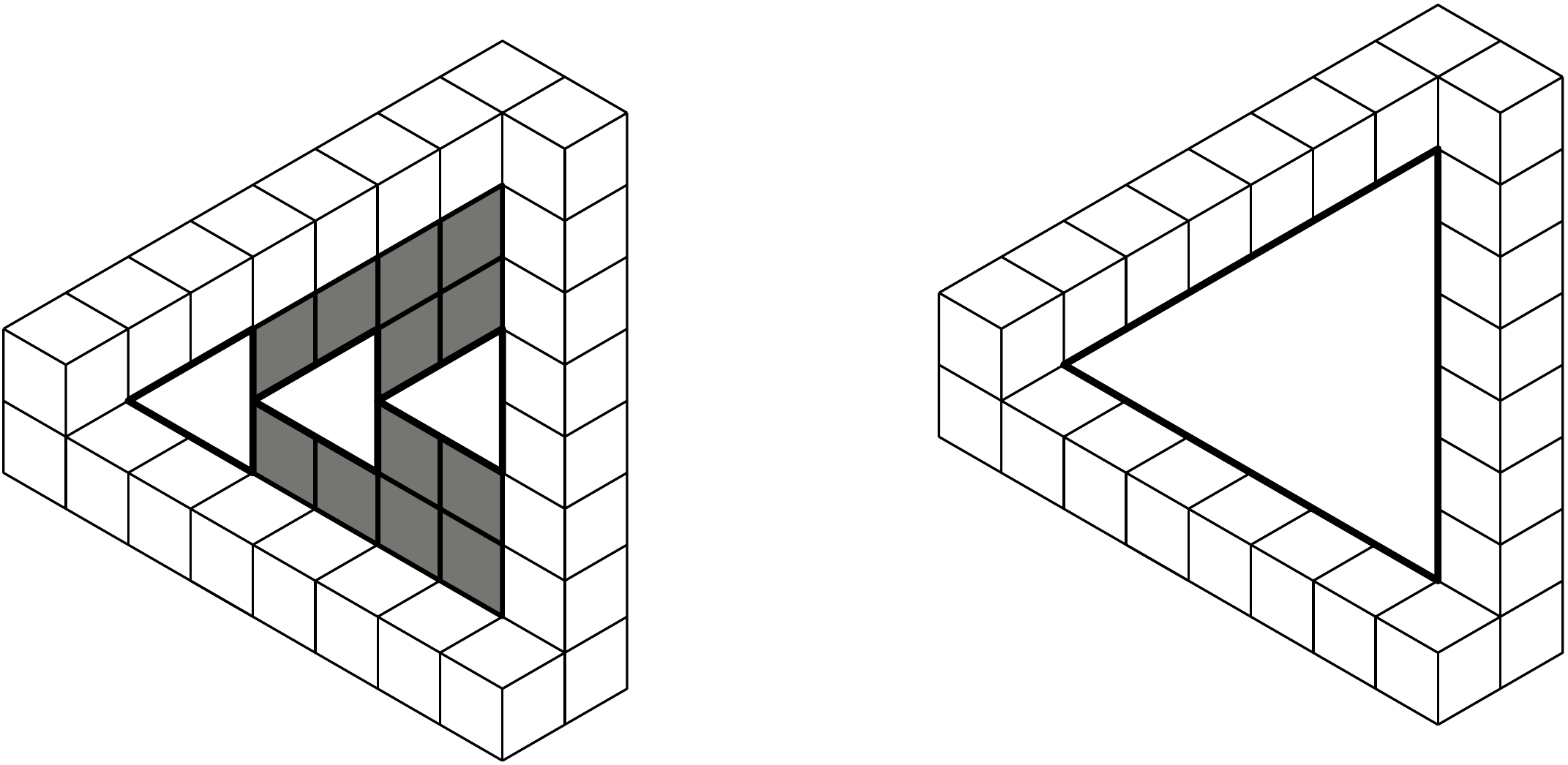}\caption{A set of contiguous triangular 
holes 
of side length two where the forced unit rhombi are coloured grey, left, and 
the larger induced hole of side length six, right.}\label{fig:Esch}
\end{figure}
In order to prove Theorem~\ref{thm:MainThm} an exact formula 
(Theorem~\ref{thm:ExactFull}) is established in Section~\ref{sec:Exact} that 
counts rhombus tilings of a \emph{holey 
hexagon}\footnote{This term was first coined by Propp in~\cite{ProppHoley} to 
describe a hexagonal region that contains a set of holes in its interior.} 
centred at some origin, $O$, with sides of length $n,2m,n,n,2m,n$ (going 
clockwise from the southwest side), containing $p$-many left and ($p$-many) 
right 
pointing horizontally collinear triangular holes of side length two lying along 
the horizontal line that intersects the origin. Such a region is denoted 
$\Hreg{n}{2m}{R}{L}$, where $R=\{r_1,\dots,r_p\}$ and $L=\{l_1,\dots,l_p\}$ are 
sets of unique integers that correspond to lattice distances of the midpoint of 
the vertical sides of the right and left pointing holes (respectively) from 
$O$. An example of a holey hexagon may be found in Figure~\ref{fig:FullTile}.
\begin{rmk}\label{Remark:Contig}
A string of $k$-many contiguous\footnote{A set of horizontally collinear holes 
are \emph{contiguous} if no horizontal rhombi can fit between them.} 
triangular holes of side length two 
is equivalent to a triangular hole of side length $2k$, since dimers 
are forced within the ``folds" of the holes (see Figure~\ref{fig:Esch}), thus 
inducing a larger hole. It is therefore sufficient to consider holey hexagons 
containing holes of side length two, and holes of a larger even side length 
shall be referred to as \emph{induced holes}.
\end{rmk}

\begin{figure}[t]\begin{center}
\includegraphics[scale=0.40]{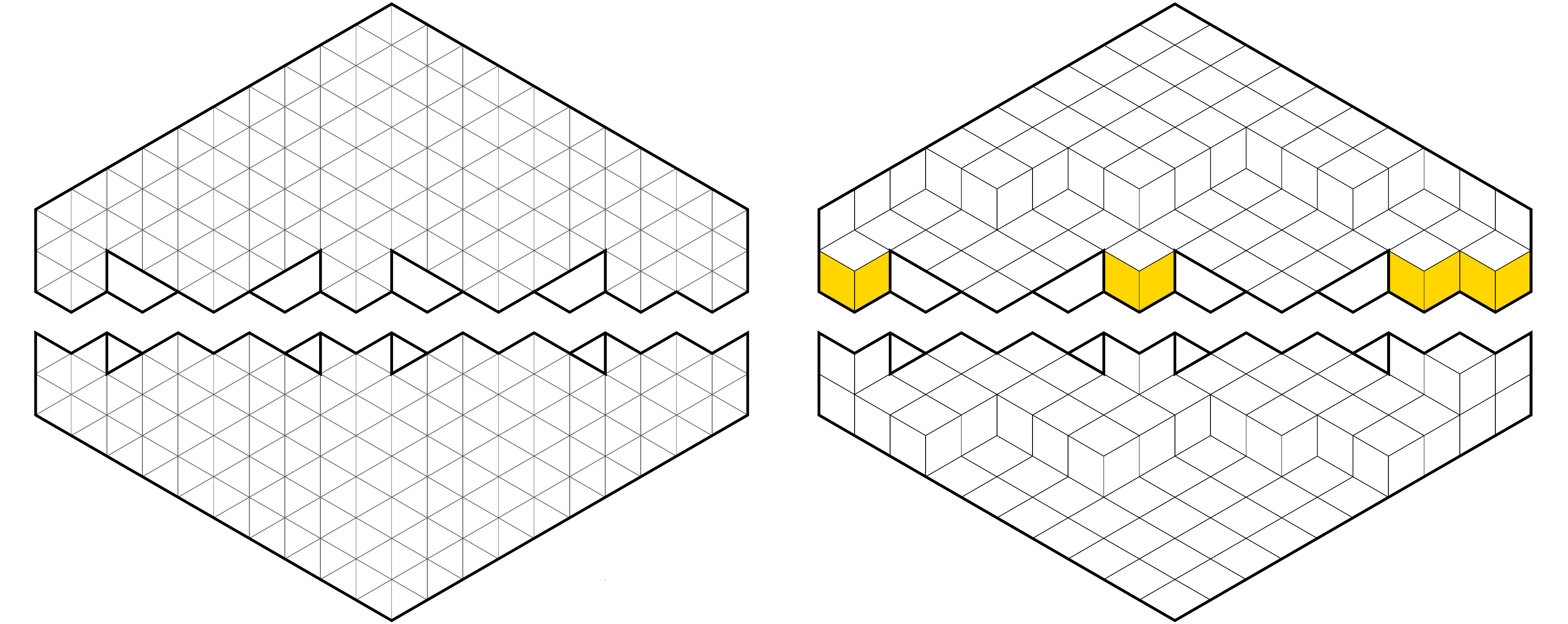}\caption{The regions 
\HPreg{10}{4}{\{0,-8\}}{\{-2,6\}}, upper left, and 
\HMreg{10}{4}{\{0,-8\}}{\{-2,6\}}, lower left, together with tilings of each 
region, right, where the pairs of yellow tiles have a combined weight of 
$2$.}\label{fig:HalfTile}\end{center}
\end{figure}

Theorem~\ref{thm:ExactFull} follows from splitting \Hreg{n}{2m}{R}{L} into two 
subregions, each obtained by cutting along the zig-zag line that proceeds just 
below the horizontal line intersecting the origin. The upper region is denoted 
\HPreg{n}{2m}{R}{L}, the lower \HMreg{n}{2m}{R}{L}. According to Ciucu's 
factorisation theorem~\cite{CiucuFact},
\begin{equation}\label{eq:Fact}M(\Hreg{n}{2m}{R}{L})=M(\HMreg{n}{2m}{R}{L}
)\cdot M_w(\HPreg{n}{2m}{R}{L}),\end{equation}
where $M_w(\HPreg{n}{2m}{R}{L})$ denotes the weighted count of tilings of 
\HPreg{n}{2m}{R}{L}, where every pair of unit rhombi that lie within the 
``folds" of the lower zig-zag 
boundary have a combined weight of $2$ (see 
Figure~\ref{fig:HalfTile}).

Exact enumerative formulas (Theorem~\ref{thm:HMthm} and 
Theorem~\ref{thm:HPthm}) 
that count (weighted) tilings of these subregions follow from
translating tilings to families of non-intersecting lattice paths in the usual 
way. According to~\cite{LGV}, enumerating such families of paths amounts to 
evaluating two determinants and 
the product of these determinant evaluations then yields 
Theorem~\ref{thm:ExactFull} by way of Ciucu's factorisation 
result~\eqref{eq:Fact}.

\begin{figure}
\includegraphics[scale=0.5]{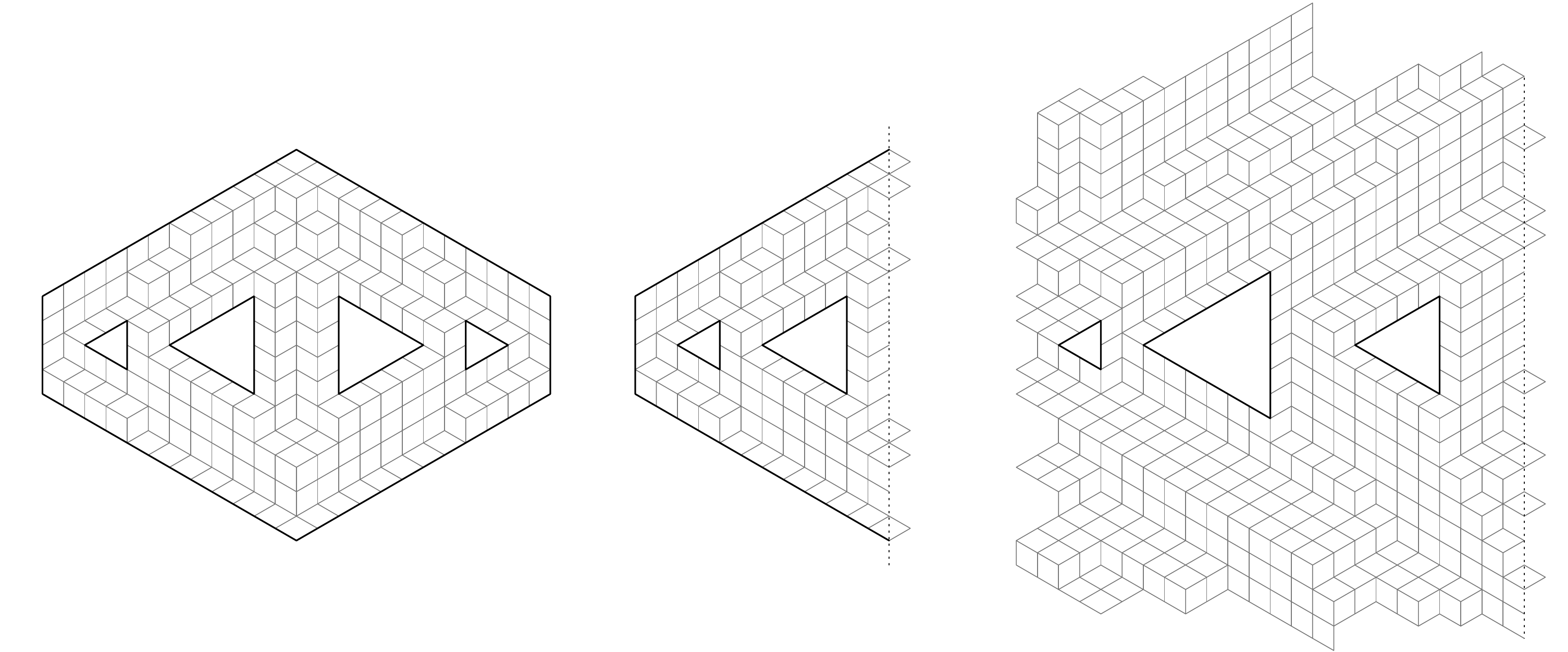}
\caption{A vertically symmetric tiling of the holey hexagon \Hreg{12}{4}{R}{L} 
where $L=\{-8,-4,-2\}$ and $R=\{2,4,8\}$, left, together with the corresponding 
tiling of 
\Vreg{12}{4}{L}, centre, and a set of left pointing holes in a sea of dimers 
constrained on the right by a free boundary, right.}
\label{fig:Vert}
\end{figure}

Enumerating tilings of \Hreg{n}{2m}{R}{L} in this way 
also gives, for free, two enumeration formulas for certain symmetry classes of 
tilings of \Hreg{n}{2m}{R}{L}. It should be clear that tilings of 
\HMreg{n}{2m}{R}{L} correspond to horizontally symmetric tilings of 
\Hreg{n}{2m}{R}{L}. Moreover, if each $r$ in $R$ is positive and satisfies 
$r=-l$ for some $l$ in $L$, then according to Ciucu and 
Krattenthaler~\cite{KrattFact} the weighted count of 
tilings of the upper region, $M_w(\HPreg{n}{2m}{R}{L})$, is equal to the number 
of vertically symmetric tilings of \Hreg{n}{2m}{R}{L}, which correspond to 
tilings of the left half of \Hreg{n}{2m}{R}{L} constrained on the right by a 
vertical free boundary that intersects the origin (here a boundary is 
considered 
free if unit rhombi are permitted to protrude halfway across it). Such a region 
shall be denoted \Vreg{n}{2m}{L} and an example of a tiling of such a region 
may be found in Figure~\ref{fig:Vert}.

Tilings of \Vreg{n}{2m}{L} correspond to tilings of the left half of the plane 
constrained on the right by a vertical free boundary as $n$ and $m$ are sent 
to infinity. The correlation function of holes that lie within tilings of this 
half plane may be 
interpreted as the interaction between a set of left pointing holes and a 
vertical free boundary (see Figure~\ref{fig:Vert}, right). Ciucu and 
Krattenthaler~\cite{CiuKrattInt} considered such an interaction for a single 
left pointing hole of side length two, which is a special case of the following 
theorem.

\begin{figure}\centering
\includegraphics[scale=0.4]{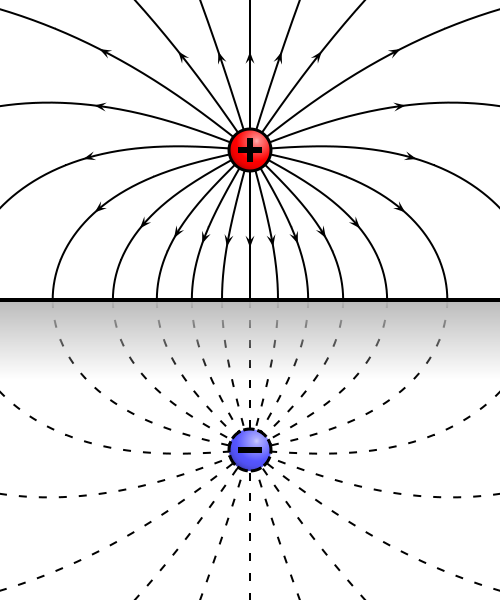}\caption{A diagram displaying the method 
of 
images, taken from~\cite{Wiki}. The electric field induced by the positive 
charge, $p$, near the horizontal straight line conductor is the same as the 
electric field induced by the $p$ and its imaginary charge $\hat{p}$ obtained 
by reflecting $p$ through the conductor.}
\label{fig:MethImg} 
\end{figure}

\begin{thm}\label{thm:VertCor}
The interaction between a set of horizontally collinear left pointing holes of 
any even side length 
in a sea of unit 
rhombi and a right vertical free boundary is
$$\omega_V(\mathcal{H})\thicksim\prod_{h\in\mathcal{H}}K_h\prod_{
1\le j< 
i\le |\mathcal{H}|}d(h_i,h_j)^{\frac{1}{4}q(h_i)q(h_j)},$$
where $\mathcal{H}$ indexes both the left pointing holes and their reflections 
in the free boundary, and 
$$K_h=\prod_{s=0}^{\frac{1}{2}|q(h)|-1}\frac{3^{s/2}\Gamma(s+1)}{\sqrt{2\pi}}.$$
\end{thm}
\begin{rmk}
The above result is in fact the square root of the result given in 
Theorem~\ref{thm:MainThm}, and indeed this relationship is analogous to 
well-known and established physical laws. Consider a point charge $p$ with 
signed magnitude situated near a straight line equipotential conductor. The 
method of images~\cite[Chapter 6]{Feyn} states that in order to calculate the 
electrostatic energy of the system one may replace the straight line conductor 
with an imaginary charge, $\hat{p}$, of opposite signed magnitude to that of 
$p$ situated at the position specified by reflecting $p$ through the conductor 
since the electric field induced by the charge(s) is the same for both 
arrangements (see Figure~\ref{fig:MethImg} for an illustration of this 
principle). The electrostatic energy of the system with the conductor is then 
one half of the electrostatic energy of the system consisting of $p$ and its 
imaginary counter-charge $\hat{p}$. Thus the result above may be obtained, up to 
a multiplicative constant, by taking the exponential of the negative of half 
the electrostatic energy of the the system obtained by considering the set of 
left pointing holes described above together with their reflections through the 
vertical free boundary as a set of point charges with signed magnitude given by 
$q$. One should see that this is equivalent (again up to some multiplicative 
constant) to taking the square root of the result given in 
Theorem~\ref{thm:MainThm}.
\end{rmk}
It should be noted that Theorem~\ref{thm:MainThm} and Theorem~\ref{thm:VertCor} 
are specialistions of more general results that appear in 
Section~\ref{sec:Asymp}, where the 
sides of the holey hexagon $\Hreg{n}{2m}{R}{L}$ may approach infinity at 
different rates (that is, $2m\thicksim\xi n$ for some real positive $\xi$). For 
$\xi\neq 1$ these interactions either blow up or shrink exponentially, thus the 
results presented above are for the special case where $\xi=1$. While this 
special case agrees completely with Ciucu's tori result~\cite{CiuScale}, the 
asymptotic analysis presented in Section~\ref{sec:Asymp} shows 
that in the planar case such an interaction depends somewhat delicately on the 
rate at which the ``edges" of the region approach infinity. A discussion of 
similar findings may also be found in~\cite[Remark 1]{CiuKrattInt}.

The following section establishes exact formulas that count weighted tilings 
of \HPreg{n}{2m}{R}{L} and \HMreg{n}{2m}{R}{L}. The asymptotic behaviours of 
these formulas are established in Section~\ref{sec:Asymp}. Throughout both 
sections the following notation for generalised hypergeometric series is used:
$$\pFq{p}{q}{a_1,\dots, 
a_p}{b_1,\dots,b_q}{z}=\sum_{k=0}^\infty\frac{(a_1)_k\cdots(a_p)_k}{
(b_1)_k\cdots (b_q)_k}\frac{z^k}{k!},$$
where $(\alpha)_\beta$ is the Pochhammer symbol, defined to be
$$(\alpha)_\beta=\begin{cases} \alpha 
\cdot(\alpha+1)\cdots(\alpha+\beta-1)&\beta>0,\\
1&\beta=0.\end{cases}$$
\section{An Exact Formula}\label{sec:Exact}
The main goal of this section is to prove the following enumerative 
formula.
\begin{thm}\label{thm:ExactFull}
The number of tilings of \Hreg{n}{2m}{R}{L} is
$$\left(\prod_{i=1}^{n}\prod_{j=1}^{2m}\prod_{k=1}^n\frac{i+j+k-1}{i+j+k-2}
\right)\cdot\det \widecheck{E}_{R,L}\cdot\det \widehat{E}_{R,L},$$
where $\widecheck{E}_{R,L}$ and $\widehat{E}_{R,L}$ are $p\times p$ matrices 
defined in 
Theorems~\ref{thm:HMthm} and~\ref{thm:HPthm} below.
\end{thm}
\begin{rmk}
The leftmost product in the above theorem is easily recognisable as MacMahon's 
much celebrated box formula~\cite{CombAnal}, named so because it counts the 
number of unit cube representations of plane partitions that fit inside an 
$n\times2m\times n$ box. This is 
equivalent to the number of rhombus tilings of an \emph{un-holey 
hexagon}\footnote{A hexagon that contains no holes.} of side lengths 
$n,2m,n,n,2m,n$ (going clockwise from the southwest edge) and is denoted 
$H_{n}{2m}$.
\end{rmk}
The above theorem follows from establishing exact formulas that enumerate the 
(weighted) number of tilings of the regions \HPreg{n}{2m}{R}{L} and 
\HMreg{n}{2m}{R}{L}. According to Ciucu's 
factorisation theorem~\eqref{eq:Fact}, the product of these formulas 
counts the total number of tilings of \Hreg{n}{2m}{R}{L}.
\begin{figure}[t]\begin{center}
\includegraphics[scale=0.40]{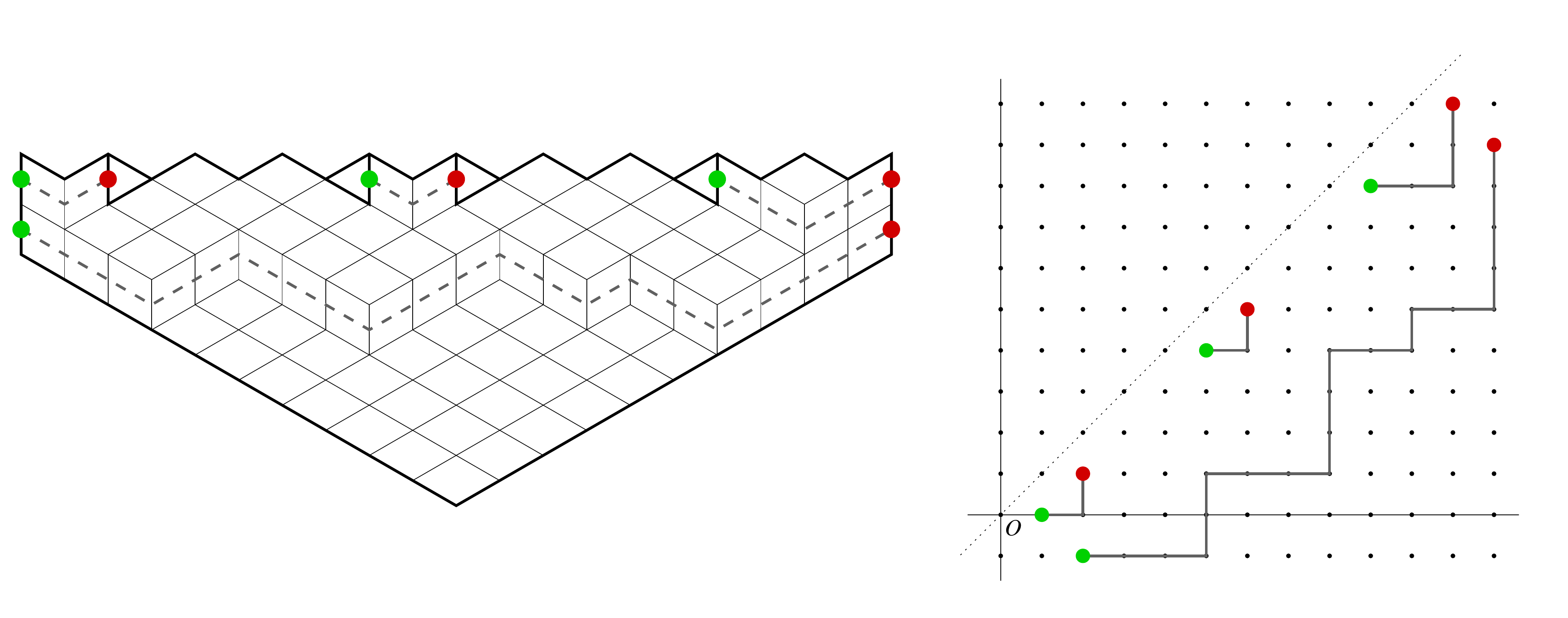}\caption{A 
tiling of \HMreg{10}{4}{\{0,-8\}}{\{-2,6\}} 
displaying lattice paths across 
unit rhombi, left, and the corresponding set of 
lattice paths starting at a set of green points 
and ending at a set of red ones, right.}\label{fig:HMlat}\end{center}
\end{figure}
To begin, one translates tilings of these 
regions into families of lattice paths across dimers in the usual 
way\footnote{To generate such a family, start points are set in the centre of 
the vertical edges of dimers that lie along the west edge of each region, along 
with the dimers that lie along the vertical edge of each left pointing 
triangular hole. One then draws a path across dimers by travelling from one 
side 
of a dimer to the opposite parallel side, thereby creating a family of 
non-intersecting paths that corresponds to precisely one tiling.}, 
which 
in turn correspond to families of non-intersecting lattice paths consisting of 
north and east steps that begin at a set of points 
$A=(A_1,\dots,A_{m+p})$ and end at a set of points $E=(E_1,\dots,E_{m+p})$, 
where
$$A_i=\begin{cases}(i,1-i)&1\le i\le m,\\
(\tfrac{n}{2}+\frac{l_{i-m}}{2}+1,\tfrac{n}{2}+\frac{l_{i-m}}{2})&m+1\le i\le 
m+p,\end{cases}$$
and
$$E_j=\begin{cases}(n+j,n+1-j)&1\le j\le m,\\
(\tfrac{n}{2}+\frac{r_{j-m}}{2}+1,\tfrac{n}{2}+\frac{r_{j-m}}{2})&m+1\le j\le 
m+p.\end{cases}$$

\begin{figure}[t]\begin{center}
\includegraphics[scale=0.40]{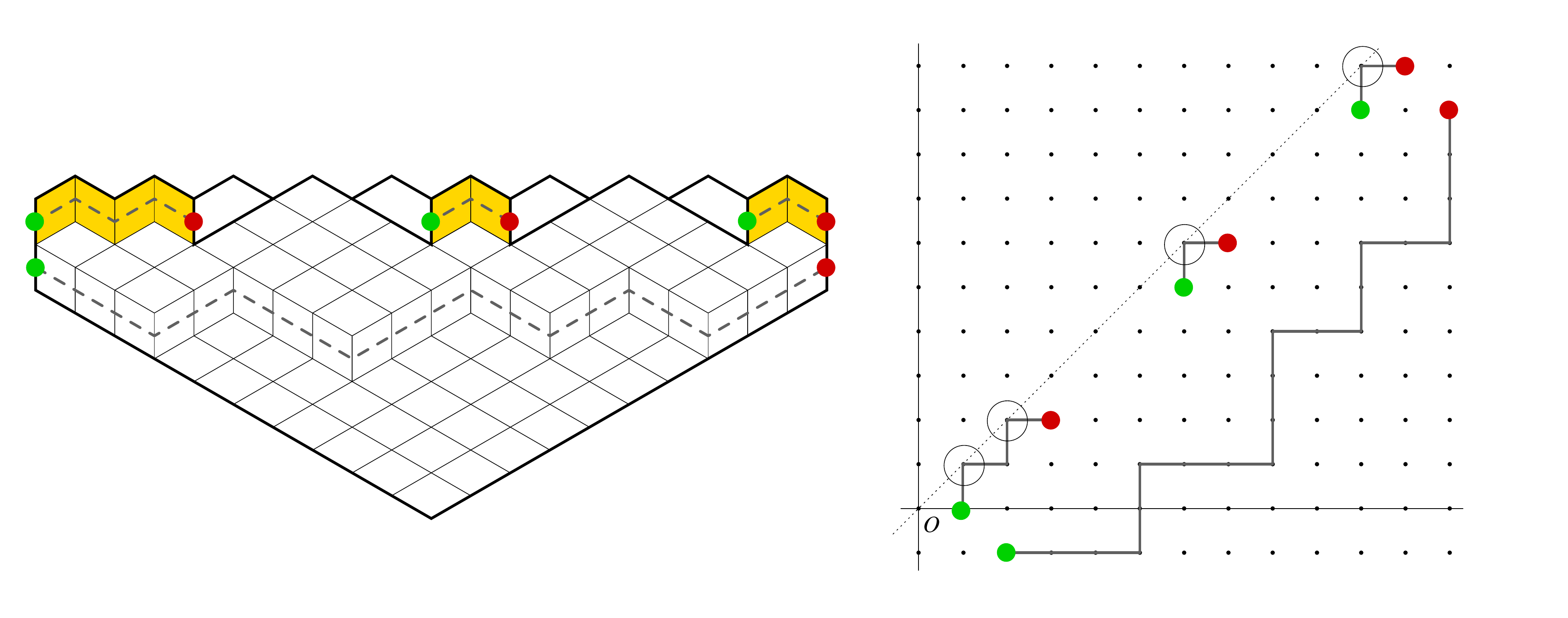}\end{center}\caption{A 
tiling of \HPreg{10}{4}{\{0,-8\}}{\{-2,6\}} 
displaying lattice paths across 
unit rhombi, left, and the corresponding set of  
lattice paths starting at a set of green points 
and ending at a set of red ones, right, where the 
yellow weighted tiles on the left correspond to 
the circled points the touch the main diagonal on 
the right.}\label{fig:HPlat}
\end{figure}

Tilings of \HMreg{n}{2m}{R}{L} correspond to families of non-intersecting 
lattice paths that begin at $A$ and end at $E$ such that no path touches the 
main diagonal (that is, the line $y=x$), see Figure~\ref{fig:HMlat}.
The weighted count of tilings of 
\HPreg{n}{2m}{R}{L} instead correspond to the weighted count of families of 
non-intersecting paths from $A$ to $E$ that do not extend above the main 
diagonal, and where each path $P$ that touches the main diagonal at $T(P)$ many 
points has a weight of $2^{T(P)}$ (see Figure~\ref{fig:HPlat}).

\begin{rmk}\label{rmk:Perm}
Suppose $\sigma$ is a permutation on $m+p$ letters that maps each start point 
$A_i$ to an end point $E_{\sigma(i)}$. It should be clear that due to the 
constraints on the families of paths described above, every family of 
non-intersecting paths arises from precisely one permutation which is 
uniquely determined by the positioning of the triangular holes relative to each 
other.
\end{rmk}
According to the well-known theorem of Lindstr\"{o}m~\cite{Lind}, Gessel, and 
Viennot~\cite{LGV}, if $\sigma$ is the only permutation on $m+p$ letters that 
gives rise to a family of non-intersecting paths that begin at $A$ and end at 
$E$, then the number of such (weighted) paths is given by 
the determinant of the matrix $P=(P_{i,j})_{1\le i,j\le m+p}$, where 
$P_{i,j}$ denotes the weighted count of the number of lattice paths from 
$A_i$ to $E_{j}$. To be more precise, suppose that every vertical or 
horizontal unit step (edge) between a point $a$ and a point $b$ has weight 
$e_w(a,b)$ (where the weights are elements of some commutative ring). Then the 
weight of a path $\mathcal{P}$ is the product over all the unit edges that 
comprise it, thus if 
$\mathscr{P}(A_i\rightarrow E_j)$ denotes the family of paths beginning at a 
point $A_i$ and 
ending at a point $E_j$ then
$$P_{i,j}=\sum_{\mathcal{P}\in \mathscr{P}(A_i\rightarrow E_j)}\prod_{(a,b)\in 
\mathcal{P}}e_w(a,b).$$
Note that if all edge weights are 1 then $P_{i,j}$ is simply the number 
of paths from 
$A_i$ to $E_j$.
 \begin{rmk}
 It shall be assumed from now on that the only permutation $\sigma$ that gives 
rise to a family of non-intersecting paths from $A$ to $E$ is the identity, for 
if not one may permute the labels of the start or end points in such a way that 
this 
holds. This corresponds to interchanging rows or columns in the matrix 
$P$, thus the number of such non-intersecting lattice paths from $A$ to $E$ is 
given by the absolute value of its determinant.
 \end{rmk}
 For tilings of the region \HMreg{n}{2m}{R}{L}, each path from $A_i$ to $E_j$ 
has a weight of 1, thus the number of non-intersecting lattice paths from $A$ 
to 
$E$ is given by
 $$|\det (\mathscr{P'}(A_i\rightarrow E_j))_{1\le i,j\le m+p}|,$$ 
 where the function $\mathscr{P'}((a,b)\rightarrow (c,d))$ denotes the number 
of lattice paths that start at the point $(a,b)$ and end at the point $(c,d)$ 
and never touch the main diagonal. A straightforward argument shows that the 
number of such paths is given by
 $$\mathscr{P'}((a,b)\rightarrow (c,d))=\mathscr{P}((a,b)\rightarrow 
(c,d))-\mathscr{P}((a,b)\rightarrow(d,c)),$$
 where $\mathscr{P}((a,b)\rightarrow (c,d))$ denotes the function that counts 
the number of ordinary lattice paths beginning at the point $(a,b)$ and ending 
at the point $(c,d)$. Since this function $\mathscr{P}$ is given by a binomial 
coefficient it follows that
 $$M(\HMreg{n}{2m}{R}{L})=|\det \widecheck{Q}|,$$
 where the matrix $\widecheck{Q}=(\widecheck{Q}_{i,j})_{1\le i,j\le m+p}$ has 
$(i,j)$-entries given by
 
$$\widecheck{Q}_{i,j}=\displaystyle\begin{cases}\binom{2n}{n+j-i}-\binom{2n}{
n+1-i-j } &1\le 
i,j\le m,\\  
\frac{2i-1}{n+r_{j-m}+1}\binom{n+r_{j-m}+1}{n/2+r_{j-m}/2+1-i}& 
i\in\{1,\dots,m\},j\in\{m+1,\dots,m+p\},\\ 
\frac{2j-1}{n-l_{i-m}+1}\binom{n-l_{i-m}+1}{n/2-l_{i-m}/2-1+j}&i\in\{m+1,\dots,
m+p\},j\in\{1,\dots,m\}, \\
  
\frac{1}{r_{j-m}-l_{i-m}+1}\binom{r_{j-m}-l_{i-m}+1}{r_{j-m}/2-l_{i-m}/2}&1\le 
i,j\le m+p.
 \end{cases}$$
 \begin{rmk}\label{rmk:Binom}
 In the above definitions (and indeed throughout this article) the binomial 
coefficient shall be interpreted in the ``natural" way, that is, for integers 
$n$ and $k$,
 $$\binom{n}{k}=\begin{cases}\frac{\Gamma(n+1)}{\Gamma(k+1)\Gamma(n-k+1)}&0\le 
k\le n,\\
 0& \textrm{otherwise}.\end{cases}$$
 \end{rmk}
 \begin{prop}\label{prop:HMLU}
 The matrix $\widecheck{Q}$ defined above has the following $LU$-decomposition
 $$\widecheck{Q}=L\cdot U,$$
 where $L=(L_{i,j})_{1\le i,j\le m+p}$ has $(i,j)$-entries given by
 $$L_{i,j}=\begin{cases}A_n(i,j)&1\le j< i\le m,\\
				   B_{n}(i,j) &m+1\le i\le m+p,1\le j\le m,\\
				   E_{n,m}(i,j) & m+1\le j<i\le m+p,\\
				   1&i=j,1\le j\le m+p,\\
				   0& otherwise,
				   \end{cases}$$
and $U=(U_{i,j})_{1\le i,j\le m+1}$ is given by
$$U_{i,j}=\begin{cases}C_{n}(i,j)&1\le i\le j\le m,\\
				     D_{n}(i,j)&1\le i\le m, m+1\le j\le m+p,\\
				     F_{n,m}(i,j)&m+1\le j\le i\le m+p,\\
				     0 &otherwise,\end{cases}$$
such that
\begin{align*}
A_{n}(i,j)&=\frac{\Gamma(2 i) \Gamma(n+1) \Gamma(i+j-1) \Gamma(2 j+n)}{\Gamma(2 
i-1) \Gamma(2 j) \Gamma(i-j+1) 
\Gamma(j-i+n+1) \Gamma(i+j+n)},\\
B_{n}(i,j)&=\frac{(-1)^{j+1} \Gamma(j+n-1)\Gamma(2 j+n) \Gamma(n-l_{i-m}+1) 
\Gamma(j+\frac{l_{i-m}}{2}+\frac{n}{2}-1)}{2 \,\Gamma(j) \Gamma(2 j+2 n-2) 
\Gamma(\frac{n}{2}-\frac{l_{i-m}}{2}+1) 
\Gamma(\frac{l_{i-m}}{2}+\frac{n}{2}) 
\Gamma(j-\frac{l_{i-m}}{2}+\frac{n}{2}+1)},\\
C_{n}(i,j)&=\frac{\Gamma(2 j) \Gamma(n+1) \Gamma(i+j-1) 
\Gamma(2i+2n-1)}{\Gamma(2 j-1) \Gamma(j-i+1) \Gamma(2 i+n-1) 
\Gamma(i-j+n+1) \Gamma(i+j+n)},\\
D_{n}(i,j)&=\frac{(-1)^{i+1} \Gamma(2 i+1) \Gamma(i+n) \Gamma(n+r_{j-m}+1) 
\Gamma(i+\frac{n}{2}-\frac{r_{j-m}}{2}-1)}{2\, \Gamma(2 i+n-1)\Gamma(i+1) 
\Gamma(\frac{n}{2}-\frac{r_{j-m}}{2})\Gamma (\frac{n}{2}+\frac{r_{j-m}}{2}+1) 
\Gamma(i+\frac{n}{2}+\frac{r_{j-m}}{2}+1)},
\end{align*}
and $E_{n,m}(i,j)$ and $F_{n,m}(i,j)$ are functions satisfying
$$\widecheck{Q}_{i,j}=\sum_{s=1}^mB_{n}(i,s)D_{n}(s,j)+\sum_{s=m+1}^{\min(i,j)}
E_ { n , m } (i ,
s) F_{n,m}(s,j)$$
for $m+1\le i,j\le m+p$.
 \end{prop}
 \begin{proof}
 The proof of the above proposition is very similar to that of Theorem~5.5 
in~\cite{TG1} in that it relies on showing that the following identities hold:
 \begin{enumerate}[(i)]
 \item\label{HMLU:P1} $\sum_{s=1}^{\min (i,j)}A_{n}(i,s) 
C_{n}(s,j)=\binom{2n}{n+j-i}-\binom{2n}{n+i-j}$;
 \item\label{HMLU:P2}$\sum_{s=1}^{i}A_n(i,s) 
D_n(s,j)=\frac{2i-1}{n+r_{j-m}+1}\binom{n+r_{j-m}+1}{n/2+r_{j-m}/2+1-i}$;
 \item\label{HMLU:P3}$\sum_{s=1}^jB_n(i,s) 
C_n(s,j)=\frac{2j-1}{n-l_{i-m}+1}\binom{n-l_{i-m}+1}{n/2-l_{i-m}/2+1-j}$.
 \end{enumerate}
 Case~\eqref{HMLU:P1} follows immediately from the observation that $A_n(i,j)$ 
and $C_n(i,j)$ defined above are equal to $A_n'(i,j)$ and $C'_n(i,j)$ 
(respectively) in Theorem~5.5 of the aforementioned article. Similarly, by 
replacing $k$ with $-r_{j-m}$ in the proof of case~(ii) of that same theorem 
one 
sees that case~\eqref{HMLU:P2} above also holds. 
 
The proof of the third case is in much the same vein, since the left hand side 
satisfies the following recurrence
 \begin{multline*}
 (2j+1)(2j+l_{i-m}-n-2)\sum_{s=1}^jB_n(i,s) 
C_n(s,j)\\+(2j-1)(2j+l_{i-m}-n-2)\sum_{s=1}^{j+1}B_n(i,s) C_n(s,j+1)=0,
 \end{multline*}
 and one may easily check that the above equation holds when the sums are 
replaced with the corresponding expression from the right hand side of 
case~\eqref{HMLU:P3}. Verifying this expression holds for initial conditions 
completes the proof.
 \end{proof}
 
\begin{rmk}
 The third case follows directly from the observation that $B_n(i,s) 
C_n(s,j)$ is equal to $A_n(j,s) D_{n}(s,i)$ with $r_{j-m}$ replaced by 
$-l_{i-m}$. The proof is presented above in such a way as to give a flavour of 
the overall approach required to prove many of the results in~\cite{TG1}, 
some of which 
are indeed specialisations of the results presented here.
 \end{rmk}
 
 \begin{thm}\label{thm:HMthm}
 The number of rhombus tilings of \HMreg{n}{2m}{R}{L} is
$$\left(\binom{n+m-1}{n-1}\prod_{i=1}^{n-2}\prod_{j=i}^{n-2}\frac{2m+i+j+1}{
i+j+1}\right)\cdot|\det \widecheck{E}_{R,L}|,$$
where $\widecheck{E}_{R,L}$ is the $p\times p$ matrix with $(i,j)$-entries 
given by
\begin{multline*}\check{e}_{i,j}=\pFq{4}{3}{\tfrac{r}{2}-\tfrac{n}{2}+1,1,\tfrac
{r}{2}-\tfrac{l}{2}+2,\tfrac{n}{2}+\tfrac{r}{2}+\tfrac{1}{2}}{m+\tfrac{n}{2}+
\tfrac{r}{2}+2,\tfrac{r}{2}-m-\tfrac{n}{2}+2,\tfrac{r}{2}-\tfrac{l}{2}+\tfrac{3}
{2} }{1}\\\times\frac { \Gamma 
(m+n+1) \Gamma 
(\frac{n}{2}+\frac{r_j}{2}+\frac{1}{2}) \Gamma (\frac{l_i}{2}+ m+\frac{n}{2}) 
\Gamma 
(m+\frac{n}{2}-\frac{r_j}{2}-1)}{\Gamma 
(\frac{n}{2}-\frac{r_j}{2}) \Gamma (m-\frac{l_i}{2}+\frac{n}{2}+1) 
\Gamma ( 
m+\frac{n}{2}+\frac{r_j}{2}+2))}\\\times\frac{2^{r_j-l_i+2} \Gamma 
(m+\frac{3}{2}) 
\Gamma (\frac{n}{2} -\frac{l_i}{2}+\frac{1}{2}) }{\pi  (r_j-l_i+1) \Gamma (m) 
\Gamma 
(\frac{l_i}{2}+\frac{n}{2}) \Gamma (m+n-\frac{1}{2})}
\end{multline*}
if $r_j>l_i$ and
\begin{multline*}
\check{e}_{i,j}=-\frac{2^{-l_i+r_j+2} \Gamma (m+\tfrac{3}{2}) \Gamma 
(\tfrac{1}{2} (-l_i+n+1)) \Gamma (m+n+1) \Gamma (\tfrac{1}{2} 
(n+r_j+1))}{3 \pi  \Gamma (m) \Gamma (\frac{1}{2} (-l_i+n+4)) 
\Gamma (m+n-\tfrac{1}{2}) \Gamma (\frac{1}{2} 
(n+r_j+4))}
\\\times\pFq{4}{3}{2-\tfrac{l_i}{2}+\tfrac{r_j}{2},\tfrac{3}{2},m+n+1,
1-m } { 
\tfrac{n}{2}+2-\tfrac{l_i}{2},\tfrac{n}{2}+\tfrac{r_j}{2}+2,\tfrac{5}{2}}{1} 
\end{multline*}
otherwise.
 \end{thm}
 \begin{proof}
 The $LU$-decomposition in Proposition~\ref{prop:HMLU} is unique since all 
entries on the diagonal of $L$ are 1. Thus the determinant of the matrix 
$\widecheck{Q}$ is
$$\prod_{s=1}^{m+p}U_{s,s}=\left(\prod_{s=1}^m 
C_{n}(s,s)\right)\cdot\prod_{t=m+1}^{m+p}U_{t,t}.$$
The product of $C_{n}(s,s)$ over $s$ above is equal to the 
well-known formula due to Proctor~\cite{Proct} that counts transpose 
complementary plane partitions in an $n\times2m\times n$ box (equivalently, 
horizontally symmetric tilings of $H_{n,2m}$),
$$\binom{n+m-1}{n-1}\prod_{i=1}^{n-2}\prod_{j=i}^{n-2}\frac{2m+i+j+1}
{i+j+1}.$$
Although the entries of $U_{t,t}$ have not been explicitly established for 
$1\le t\le m+p$, it should be clear that their product is equal to the 
determinant of a certain $p\times p$ matrix 
$\widecheck{E}_{R,L}=(\check{e}_{i,j})_{1\le i,j\le p}$
with entries given by
\begin{align*} \check{e}_{i,j}&=\sum_{s=m+1}^{\min(i,j)}E_{n,m}(m+i,s) 
F_{n,m}(s,m+j)\\&=\widecheck{Q}_{m+i,m+j}-\sum_{s=1}^mB_{n}(m+i,s)D_{n}(s,m+j).
\end{align*}
Suppose first that $r_j>l_i$. Then the entry $\check{e}_{i,j}$ may be 
re-written as
\begin{multline*} 
\frac{1}{r_{j-m}-l_{i-m}+1}\binom{r_{j-m}-l_{i-m}+1}{r_{j-m}/2-l_{i-m}/2}-\sum_{
s=0}^{\infty}B_{n}(m+i,s+1)D_{n}(s+1,m+j)\\+\sum_{t=m}^{\infty}B_{n}(m+i,t+1)D_{
n } (t+1 , m+j).
\end{multline*}
The above sum over $s$ may be expressed as the following hypergeometric series
\begin{multline}\label{hmproof:hyp1}\frac{\Gamma(n) \Gamma(n+2)\Gamma(n-l+1) 
\Gamma(n+r+1)}{2 \Gamma(2 
n) \Gamma(\frac{n}{2}-\frac{l}{2}+1) 
\Gamma(\frac{n}{2}-\frac{l}{2}+2) \Gamma(\frac{n}{2}+\frac{r}{2}+1) 
\Gamma(\frac{n}{2}+\frac{r}{2}+2)}\\\times\pFq{5}{4}{n+1, 
\frac{n}{2}+\frac{3}{2}, \frac{3}{2}, 
\frac{l}{2}+\frac{n}{2}, \frac{n}{2}-\frac{r}{2}}{\frac{n}{2}+\frac{1}{2}, 
n+\frac{1}{2}, \frac{n}{2}-\frac{l}{2}+2, 
\frac{n}{2}+\frac{r}{2}+2}{1},\end{multline}
(note the abuse of notation: $r_j$ and $l_i$ have been replaced with $r$ and 
$l$ 
respectively). According to Slater~\cite[Appendix III.12]{Slat} this 
hypergeometric series 
satisfies the following summation formula
\begin{multline*}\pFq{5}{4}{a,\frac{a}{2}+1,b,c,d}{\frac{a}{2},a-b+1,a-c+1,a-d+1
} { 1 } \\=\frac {
\Gamma { (a-b+1) } \Gamma {( a-c+1) } 
\Gamma{(a-d+1)}\Gamma{(a-b-c-d+1)}}{\Gamma{(a+1)}\Gamma{(a-b-c+1)}\Gamma{
(a-b-d+1) } \Gamma {(a-c-d+1)}}.\end{multline*}
Applying this formula to~\eqref{hmproof:hyp1} yields
$$\frac{\Gamma(n) \Gamma(n-\frac{1}{2}+1) \Gamma(n-l+1) \Gamma(\frac{r}{2} 
-\frac{l}{2}+\frac{1}{2}) \Gamma(n+r+1)}{2 \Gamma(2 n) \Gamma(\frac{n}{2} 
-\frac{l}{2}+\frac{1}{2}) 
\Gamma(\frac{n}{2}-\frac{l}{2}+1) \Gamma(\frac{r}{2} -\frac{l}{2}+2) 
\Gamma(\frac{n}{2} +\frac{r}{2}+\frac{1}{2}) 
\Gamma(\frac{n}{2}+\frac{r}{2}+1)},$$
which may easily be shown to equal
$$\frac{1}{r-l+1}\binom{r-l+1}{r/2-l/2},$$
hence for $r_j>l_i$ the entries $\check{e}_{i,j}$ are given by
$$\sum_{t=0}^{\infty}B_{n}(m+i,t+m+1)D_{
n } (t+m+1 , m+j).$$

Prolonging the aforementioned abuse of notation, the 
above sum may be written as the following 
hypergeometric series
\begin{multline}\label{hmproof:final}\pFq{6}{5}{m+\tfrac{n}{2}+\tfrac{3}{2},
 m+\tfrac{3}{2}, \tfrac{l}{2}+m+\tfrac{n}{2}, m+n+1, 
m+\tfrac{n}{2}-\tfrac{r}{2}, 1}{m+\tfrac{n}{2}+\tfrac{1}{2}, m+n+\tfrac{1}{2}, 
m-\tfrac{l}{2}+\tfrac{n}{2}+2, 
m+1, m+\tfrac{n}{2}+\tfrac{r}{2}+2}{1}\\\times\frac{\Gamma(2 m+3) 
\Gamma(n-l+1) \Gamma(m+n) 
\Gamma(m+n+1) \Gamma(2 m+n+2) }{4\, \Gamma (m+1) \Gamma(m+2) 
\Gamma(\frac{n}{2}-\frac{l}{2}+1) 
\Gamma(\frac{l}{2}+\frac{n}{2}) \Gamma(2 m+n+1) \Gamma(2 m+2 n) 
}\\\times\frac{\Gamma(n+r+1) 
\Gamma(\frac{l}{2}+m+\frac{n}{2}) 
\Gamma(m+\frac{n}{2}-\frac{r}{2})}{\Gamma(\frac{n}{2}-\frac{r}{2}
)\Gamma(\frac{n}{2}+\frac{r}{2}+1) 
\Gamma(m-\frac{l}{2}+\frac{n}{2}+2) 
\Gamma(m+\frac{n}{2}+\frac{r}{2}+2)}.\end{multline}

The $_6F_5$ hypergeometric series in the above expression may be written as the 
following limit of a $_7F_6$ series
\begin{equation}\label{hmproof:lim1}\lim_{\epsilon\rightarrow0}\,\,\pFq{7}{6}{
2m+n+1+\epsilon , V } { W }{1 },\end{equation}
where $V$ and $W$ are the lists
$$(m+\tfrac{n}{2}+\tfrac{3}
{ 2 } +\tfrac { \epsilon } { 2 } ,
 m+\tfrac{3}{2}+\tfrac{\epsilon}{2}, 
\tfrac{l}{2}+m+\tfrac{n}{2}+\tfrac{\epsilon}{2},1+\tfrac{\epsilon}{2}, 
m+n+1+\tfrac{\epsilon}{2}, 
m+\tfrac{n}{2}-\tfrac{r}{2}+\tfrac{\epsilon}{2})$$ and 
$$(m+\tfrac{n}{2}+\tfrac{1}{2}+\tfrac{\epsilon}{2}, 
m+n+\tfrac{1}{2}+\tfrac{\epsilon}{2}, 
m-\tfrac{l}{2}+\tfrac{n}{2}+2+\tfrac{\epsilon}{2},
2m+n+1+\tfrac{e}{2},
m+1+\tfrac{\epsilon}{2}, 
m+\tfrac{n}{2}+\tfrac{r}{2}+2+\tfrac{\epsilon}{2})$$ respectively. Such a 
series satisfies the following transformation formula
\begin{multline*}
\pFq{7}{6}{a,\tfrac{a}{2}+1,b,c,d,e,a-e+n+1}{\tfrac{a}{2},a-b+1,a-c+1,a-d+1,a-e+
1,e-n}{1}\\=\frac{\Gamma{(a-d+1)}\Gamma{(a-c+1)}\Gamma{(a-b+1)}\Gamma{
(a-b-c-d+1) } } { 
\Gamma{(a-c-d+1)}\Gamma{(a-b-d+1)}\Gamma{(a-b-c+1)}\Gamma{(a+1)}}\\\times\pFq{4}
{ 3 } { b , c , d , -n } { a-e+1, -a+b+c+d, e-n}{1},
\end{multline*}
which may also be found in Slater's book~\cite[(4.3.6.4) reversed]{Slat}. 
Applying this 
transformation to~\eqref{hmproof:lim1}, permuting the elements in the 
hypergeometric series and letting $\epsilon$ tend to zero one 
obtains
\begin{equation}\label{hmproof:hyp2}\frac{(2 m + 2 n - 1) (2 m-l + n + 2)}{2 ( 
n-l - 1) (2 m + n + 1)}\pFq{4}{3}{1, \tfrac{l}{2}+m+\tfrac{n}{2}, 
\tfrac{r}{2}-\tfrac{n}{2}+1,m+\tfrac{3}{2}}{m+1, 
\tfrac{l}{2}-\tfrac{n}{2}+\tfrac{3}{2}, 
m+\tfrac{n}{2}+\tfrac{r}{2}+2}{1}.\end{equation}
One final transformation formula (see Slater~\cite[(4.3.5.1)]{Slat}),
\begin{multline*}\pFq{4}{3}{a, b, c, -n}{e, f, 
a+b+c-e-f-n+1}{1}\\=\pFq{4}{3}{-n, a, 
a+c-e-f-n+1, a+b-e-f-n+1}{a+b+c-e-f-n+1, a-e-n+1, 
a-f-n+1}{1}\\\times\frac{(e-a)_n(f-a)_n}{(e)_n(f)_n},\end{multline*}
applied to the $_4F_3$ series in~\eqref{hmproof:hyp2} gives
\begin{multline*}\frac{m (2 m+2 n-1) (2 m+n-l+2)}{(r-l+1) 
(2 m+n+1) (2 m+n-r-2)}\\\times\pFq{4}{3}{\tfrac{r}{2}-\tfrac{n}{2}+1, 1, 
\tfrac{r}{2}-\tfrac{l}{2}+2, 
\tfrac{n}{2}+\tfrac{r}{2}+\tfrac{1}{2}}{m+\tfrac{n}{2}+\tfrac{r}{2}+2, 
\tfrac{r}{2}-m-\frac{n}{2}+2, 
\tfrac{r}{2}-\tfrac{l}{2}+\frac{3}{2}}{1}\end{multline*}
as an equivalent expression for the $_6F_5$ series stated above. 

Suppose now that $r_j<l_i$. Then according to Remark~\ref{rmk:Binom}
$$\check{e}_{i,j}=-\sum_{s=1}^{m}B_n(m+i,s)D_n(s,m+j),$$
the right hand side of which may be expressed as the following limit of a 
hypergeometric series
\begin{equation}\label{eq:VW7F6}
\frac{\Gamma(n)\Gamma(n+2)\Gamma(n-l_i+1)\Gamma(n+r_j+1)}{2\,
\Gamma(2n)\Gamma(\tfrac 
{n}{2}-\tfrac{l_i}{2}+1)\Gamma(\tfrac{n}{2}-\tfrac{l_i}{2}+2)\Gamma(\tfrac{n}{2} 
+\tfrac{r_j}{2}+1)\Gamma(\tfrac{n}{2}+\tfrac{r_j}{2}+2)} \lim_ { \epsilon\to 
0}\left(\pFq{7}{6}{V}{W}{1}\right),
\end{equation}
where
$$V=(\epsilon + n + 1 , \tfrac{\epsilon}{2} + 
\tfrac{n}{2} + \tfrac{3}{2} , \tfrac{\epsilon}{2} + \tfrac{l_i}{2} + 
\tfrac{n}{2} 
, \tfrac{\epsilon}{2} + \tfrac{n}{
 2} - \tfrac{r_j}{2} , \tfrac{\epsilon}{2} + \tfrac{3}{2} , m + n + 1 , 1 - 
m)$$
and
$$W=(\tfrac{\epsilon}{2} + \tfrac{n}{2} + \tfrac{1}{2} , \tfrac{\epsilon}{2} - 
\tfrac{l_i}{2} + \tfrac{n}{2} + 2 , \tfrac{\epsilon}{2} + \tfrac{n}{2} + 
\tfrac{r_j}{2} + 2 , \tfrac{\epsilon}{2} +  n + \tfrac{1}{2} , \epsilon - m 
+ 1 , \epsilon + m +
   n + 1).$$
By applying the following transformation formula,
\begin{multline}\label{eq:7F6Trans}\pFq{7}{6}{a ,  \tfrac{a}{2} + 1 ,  b ,  c , 
 d ,  e 
, -n}{\tfrac{a}{2} ,  a - b + 1 ,  a - c + 1 ,  a - d + 1 ,  a - e + 
  1 , a + n + 
1}{1}\\=\frac{(a+1)_n(a-d-e+1)_n}{(a-d+1)_n(a-e+1)_n}\pFq{4}{3}{a - b - c + 1 , 
 d , e , -n}{a - b + 1 ,  a - c + 1 ,  -a + d + e - n}{1}\end{multline}
(see~\cite[(2.4.1.1), reversed]{Slat}) to~\eqref{eq:VW7F6} and letting 
$\epsilon$ tend to zero, one obtains
\begin{multline*}
\check{e}_{i,j}=-\frac{2^{r_j-l_i+2} \Gamma (m+\tfrac{3}{2}) \Gamma 
 (\tfrac{n}{2}-\tfrac{l_i}{2}+\tfrac{1}{2}) \Gamma (m+n+1) \Gamma 
(\tfrac{n}{2}+\tfrac{r_j}{2}+\tfrac{1}{2})}{3 \pi\,\Gamma (m) \Gamma 
(\frac{n}{2} -\tfrac{l_i}{2}+2) 
\Gamma (m+n-\tfrac{1}{2}) \Gamma (\frac{n}{2} 
+\tfrac{r_j}{2}+2)}
\\\times\pFq{4}{3}{2-\tfrac{l_i}{2}+\tfrac{r_j}{2},\tfrac{3}{2},m+n+1,
1-m } { 
\tfrac{n}{2}+2-\tfrac{l_i}{2},\tfrac{n}{2}+\tfrac{r_j}{2}+2,\tfrac{5}{2}}{1} 
\end{multline*}
thus determining $\check{e}_{i,j}$ completely.
 \end{proof}
What remains is to determine a formula that counts the weighted tilings of 
\HPreg{n}{2m}{R}{L}. According to Lindstr\"{o}m~\cite{Lind}, Gessel and 
Viennot~\cite{LGV}, 
the 
number of families of weighted paths that count such tilings is given by
$$|\det(\mathscr{P}''(A_i\rightarrow E_j))_{1\le i,j\le m+p}|,$$
where $\mathscr{P}''((a,b)\rightarrow(c,d))$ denotes the number of paths from 
the point $(a,b)$ to the point $(c,d)$ that do not extend above the main 
diagonal, with the added condition that each path $P$ that touches the main 
diagonal at $T(P)$ many 
points has a weight of $2^{T(P)}$. It is straightforward to show (an argument 
may be found in~\cite{KrattFact}) that
$$\mathscr{P}''((a,b)\rightarrow(c,d))=\mathscr{P}((a,b)\rightarrow(c,
d))+\mathscr{P}((a,b)\rightarrow(d,c)),$$
whence
$$M_w(\HPreg{n}{2m}{R}{L})=|\det \widehat{Q}|,$$
where $\widehat{Q}=(\widehat{Q}_{i,j})_{1\le i,j\le m+p}$ has $(i,j)$-entries 
given by
$$\widehat{Q}_{i,j}=\displaystyle\begin{cases}\binom{2n}{n+j-i}+\binom{2n}{
n+1-i-j } &1\le 
i,j\le m,\\  
\binom{n+r_{j-m}+1}{n/2+r_{j-m}/2+1-i}&i\in\{
1,\dots,m\}
 
,j\in\{m+1,\dots,m+p\},\\ 
\binom{n-l_{i-m}+1}{n/2-l_{i-m}/2+1-j}
&i\in\{m+1,\dots, 
m+p\},j\in\{1,\dots,m\}, \\
  
\binom{r_{j-m}-l_{i-m}+1}{r_{j-m}/2-l_{i-m}/2}&1\le 
i,j\le m+p.
 \end{cases}$$
 \begin{rmk}\label{rmk:Vert}
 As mentioned in the previous section, if the holes determined by the sets $R$ 
and $L$ are distributed symmetrically with respect to the horizontal \emph{and} 
vertical symmetry axes of \Hreg{n}{2m}{R}{L} (that is, all $r$ in $R$ are 
positive and satisfy $r=-l$ for some $l$ in $L$), then according 
to~\cite{KrattFact} the number of tilings of \HPreg{n}{2m}{R}{L} is equal to 
the number of tilings of \Vreg{n}{2m}{L} (that is, the number of vertically 
symmetric tilings of \Hreg{n}{2m}{R}{L}). This holds in particular for 
$L=R=\emptyset$ (in other words, vertically symmetric tilings of the un-holey 
hexagon $H_{n,2m}$). The proof of this striking result relies on applying 
elementary row and column operations to certain matrices- it would appear 
that a combinatorial interpretation of the relationship between these sets of 
tilings has yet to be realised.
\end{rmk}
 \begin{prop}\label{HPprop}
The matrix $\widehat{Q}$ defined above has $LU$-decomposition
$$\widehat{Q}=L'\cdot U',$$
where $L'=(L'_{i,j})_{1\le i,j\le m+p}$ is given by
$$L'_{i,j}=\begin{cases}A'_{n}(i,j)&1\le j< i\le m,\\
			B'_{n}(i,j)&m+1\le i\le m+p,1\le j\le m,\\
			E'_{n,m}(i,j)&m+1\le j<i\le m+p,\\
			1& i=j,1\le j\le m+p,\\
			0& otherwise,\end{cases}$$
and $U'=(U'_{i,j})_{1\le i,j\le m+p}$ is given by
$$U'_{i,j}=\begin{cases}C'_{n}(i,j)&1\le i\le j\le m,\\
			D'_{n}(i,j)&1\le i\le m,m+1\le j\le m+p,\\
			F'_{n,m}(i,j)&m+1\le i\le j\le m+p,\\
			0 &otherwise,\end{cases}$$
where
\begin{align*}
A'_{n}(i,j)&=\frac{\Gamma(n+1) \Gamma(i+j-1) \Gamma(2 j+n)}{\Gamma(2 j-1) 
\Gamma(i-j+1) \Gamma(j-i+n+1) 
\Gamma(i+j+n)},\\
B'_{n}(i,j)&=\frac{(-1)^{j+1} \Gamma(j+n) \Gamma(2 j+n) \Gamma(n-l_{i-m}+2) 
\Gamma(j+\frac{l_{i-m}}{2}+\frac{n}{2}-1)}{\Gamma(j) \Gamma(2 j+2 n) 
\Gamma(\frac{n}{2}-\frac{l_{i-m}}{2}+1) 
\Gamma(\frac{l_{i-m}}{2}+\frac{n}{2}) 
\Gamma(j-\frac{l_{i-m}}{2}+\frac{n}{2}+1)},\\
C'_{n}(i,j)&=\frac{\Gamma(n+1) \Gamma(i+j-1) \Gamma(2 i+2 n)}{\Gamma(j-i+1) 
\Gamma(2 i+n-1) \Gamma(i-j+n+1)\Gamma(i+j+n)},\\
D'_{n}(i,j)&=\frac{(-1)^{i+1} \Gamma(2 i-1) \Gamma(i+n) \Gamma(n+r_{j-m}+2) 
\Gamma(i+\frac{n}{2}-\frac{r_{j-m}}{2}-1)}{\Gamma(i) \Gamma(2 i+n-1) 
\Gamma(\frac{n}{2}-\frac{r_{j-m}}{2}) 
\Gamma(\frac{n}{2}+\frac{r_{j-m}}{2}+1) 
\Gamma(i+\frac{n}{2}+\frac{r_{j-m}}{2}+1)},
\end{align*}
and $E'_{n,m}(i,j)$ and $F'_{n,m}{i,j}$ are functions satisfying
$$\widehat{Q}_{i,j}=\sum_{s=1}^mB'_n(i,s)D'_n(s,j)+\sum_{s=m+1}^{\min(i,j)}E'_{n
,m}(i,
s)F'_{n,m}(s,j).$$
\end{prop}
\begin{proof}
Once again, the proof of the above theorem may be reduced to proving three 
identities, chiefly:
\begin{enumerate}[(i)]
\item $\sum_{s=1}^{\min(i,j)}A'_n(i,s)C'_n(s,j)=\binom{2n}{n+j-i}+\binom{2n}{
n+1-i-j }$;
\item$\sum_{s=1}^jA'_n(i,s)D'_n(s,j)=\binom{n-l_{i-m}+1}{n/2-l_{i-m}/2+1-j}$;
\item$\sum_{s=1}^{i}B'_n(i,s)C'_n(s,j)=\binom{n+r_{j-m}+1}{n/2+r_{j-m}/2+1-i}$.
\end{enumerate}
Proofs of these identities may also be found in~\cite[Theorem~5.2 and 
Lemma~5.3]{TG1}, since $A'_n(i,j)$, $B'_n(i,j)$, $C'_n(i,j)$, $D'_n(i,j)$ above 
are equal to $A_n(i,j)$, $B_{n,l_{i-m}}(j)$, $C_n(i,j)$, $E_{n,-r_{j-m}}(i)$ 
(respectively) from that same article.
\end{proof}
\begin{thm}\label{thm:HPthm}
The weighted count of rhombus tilings of \HPreg{n}{2m}{R}{L} is given by
$$\left(\prod_{i=1}^{n}\frac{2i+2m-1}{2i-1}\prod_ { 1\le i<j\le 
n}\frac{i+j+2m-1}{i+j-1}\right)\cdot|\det \widehat{E}_{R,L}|,$$
where $\widehat{E}_{R,L}$ is the $p\times p$ matrix with $(i,j)$-entries given 
by
\begin{multline*}
\hat{e}_{i,j}=\pFq{4}{3}{\tfrac{r_j}{2}-\tfrac{n}{2}+1,1,\tfrac{r_j}{2}- 
\tfrac{l_i}{2}+2,\tfrac{n}{2}+\tfrac{r_j}{2}+\tfrac{3}{2}}{m+\tfrac{n}{2}+\tfrac
{r_j}{2}+2,\tfrac{r_j}{2}-m-\tfrac{n}{2}+2,\tfrac{r_j}{2}-\tfrac{l_i}{2}+\tfrac{
5}{2}}{1}\\\times\frac { \Gamma (m+n+1) \Gamma 
(\tfrac{n}{2}+\tfrac{r_j}{2}+\tfrac{3}{2}) \Gamma (\tfrac{l_i}{2}+ 
\tfrac{m}{2}+\tfrac{n}{2}+1) \Gamma (m+\tfrac{n}{2}-\tfrac{r_j}{2}-1)}{ \Gamma 
(\tfrac{n}{2}-\tfrac{r_j}{2}) \Gamma (\tfrac{n}{2}-\tfrac{l_i}{2}+m+1) \Gamma 
(\tfrac{n}{2}+m+\tfrac{r_j}{2}+2)}\\\times\frac{2^{-l+r+2} \Gamma 
(m+\frac{1}{2}) \Gamma (\tfrac{n}{2}-\tfrac{l_i}{2}+\tfrac{3}{2}) }{\pi  
(r-l+3) \Gamma (m) \Gamma (\tfrac{l_i}{2}+\tfrac{n}{2}) \Gamma 
(m+n+\tfrac{1}{2})}
\end{multline*}
for $r_j<l_i$ and
\begin{multline*}
\hat{e}_{i,j}=-\frac{2^{r_j-l_i+2} \Gamma (m+\tfrac{1}{2}) \Gamma 
(\tfrac{n}{2}-\tfrac{l_i}{2}+\tfrac{3}{2}) \Gamma (m+n+1) \Gamma 
(\tfrac{n}{2}+\tfrac{r_j}{2}+\tfrac{3}{2})}{\pi  \Gamma (m) \Gamma 
(\tfrac{n}{2}-\tfrac{l_i}{2}+2) \Gamma (m+n+\tfrac{1}{2}) \Gamma 
(\tfrac{n}{2}+\tfrac{r_j}{2}+2)}\\\times\pFq{4}{3}{2+\tfrac{r_j}{2}-\tfrac{l_i}{
2 } 
,\tfrac{1}{2},m+n+1,1-m}{\tfrac{n}{2}-\tfrac{l_i}{2}+2,\tfrac{n}{2}+\tfrac{r_j}{ 
2}+2,\tfrac{3}{2}}{1}
\end{multline*}
otherwise.
\end{thm}
\begin{proof}
The $LU$-decomposition of $\widehat{Q}$ is unique, thus its determinant is
$$\prod_{s=1}^{m+p}U'_{s,s}=\left(\prod_{s=1}^mC'_{n}(s,s)\right)\cdot\prod_{t=
m+1}^{m+p}U'_{t,t}.$$
The product of $C'_n(s,s)$ over $s$ is in fact the number of tilings of 
the un-holey hexagon $H_{n,2m}$ that are vertically symmetric (see 
Remark~\ref{rmk:Vert}). A formula for such tilings was conjectured by 
MacMahon~\cite{CombAnal} to be
$$\prod_{i=1}^{n}\frac{2i+2m-1}{2i-1}\prod_ { 1\le i<j\le 
n}\frac{i+j+2m-1}{i+j-1},$$
which was later proved independently by both Andrews~\cite{PlanePart1} and
Gordon~\cite{GordPlane} (although Gordon's proof was published much later).

By the same argument that appears in the proof of Theorem~\ref{thm:HMthm},
$$\prod_{t=m+1}^{m+p}U'_{t,t}=\det \widehat{E}_{R,L},$$
where $\widehat{E}_{R,L}$ is a $p\times p$ matrix with $(i,j)$-entries given by
$$\hat{e}_{i,j}=\widehat{Q}_{m+i,m+j}-\sum_{s=1}^mB'_n(m+i,s)D'_n(s,m+j).$$
Precisely the same transformation and summation formulas from the proof of 
Theorem~\ref{thm:HMthm} may be applied to the right hand side of the above 
expression in order to obtain $\hat{e}_{i,j}$ as stated in the theorem.
\end{proof}

Having established Theorem~\ref{thm:HMthm} and Theorem~\ref{thm:HPthm}, 
Theorem~\ref{thm:ExactFull} follows immediately from inserting the above 
expressions for $M(\HMreg{n}{2m}{R}{L})$ and $M_w(\HPreg{n}{2m}{R}{L})$ into 
Ciucu's factorisation theorem~\eqref{eq:Fact}. Note that the absolute value of 
the determinant is no longer taken in Theorem~\ref{thm:ExactFull} since it is 
precisely the same permutation that maps start points to end points in the 
lattice path respresentations of tilings of both \HMreg{n}{2m}{R}{L} and 
\HPreg{n}{2m}{R}{L}, and so the determinants of $ \widecheck{E}_{R,L}$ and $ 
\widehat{E}_{R,L}$ have 
the same sign.

\begin{thm}\label{thm:Inj}
The absolute values of the determinants of the matrices $\widecheck{E}_{R,L}$ 
and $\widehat{E}_{R,L}$ are both bounded above by 1 for all $n,m,R,L$ where 
$-n\le r,l\le n$ for all $r\in R$ and $l\in L$.
\end{thm}

\begin{proof}
In order to prove the above theorem it is sufficient to construct a map 
$$\zeta: T_{n,2m}^{R,L}\to T_{n,2m}^{\emptyset,\emptyset}$$ that 
maps distinct tilings of $\HMreg{n}{2m}{R}{L}$ to distinct tilings of 
$\HMreg{n}{2m}{\emptyset}{\emptyset}$.

Suppose first that $|R|=|L|=1$ and consider the two possible configurations of 
holes $h_1$ and $h_2$ along the zig-zag boundary of \HMreg{n}{2m}{R}{L} (where 
$h_1$ is the leftmost unit hole, and $h_2$ the 
rightmost, see Figure~\ref{fig:PeakCon}).
\begin{figure}[t]
\includegraphics[scale=0.7]{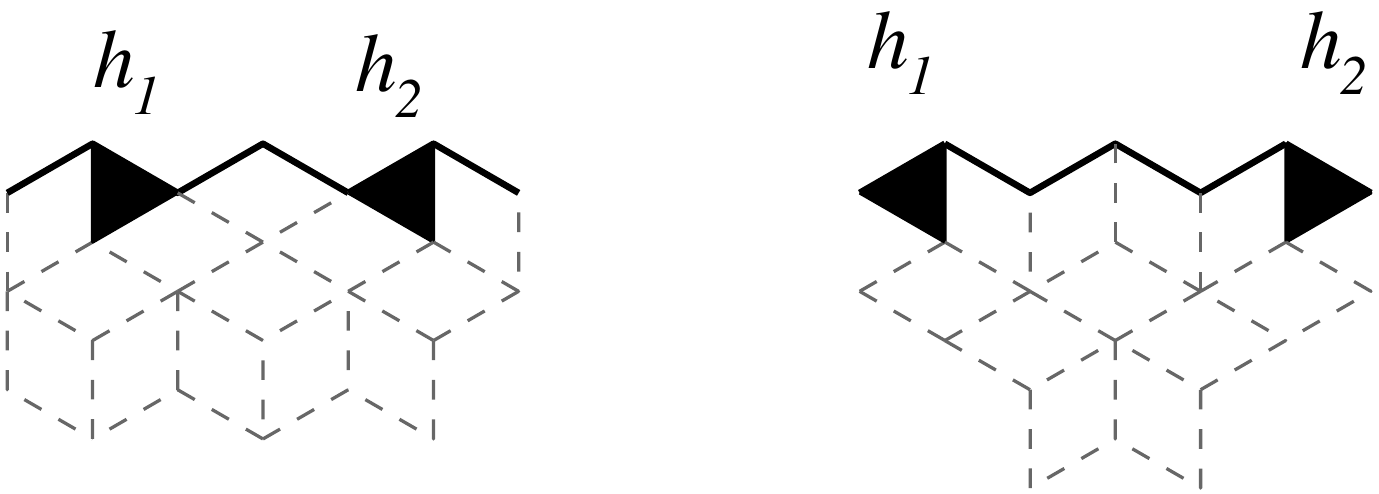}\caption{The two types 
of unit hole configuration that can occur within a 
peak along the zig-zag 
boundary.}\label{fig:PeakCon}
\end{figure}
Beginning at $h_1$, one can construct what shall be referred to as a 
\emph{propagation 
path} from $h_1$ to $h_2$ in the following 
way:
\begin{enumerate}[(i)]\item if $h_1$ is left pointing then construct a path 
across unit rhombi that begins at the midpoint of the vertical edge of $h_1$ 
and ends at the midpoint of the vertical edge of $h_2$;
\item\label{step:2} otherwise consider the path across unit rhombi that begins 
at the midpoint of the southeast edge of $h_1$ and ends at the midpoint of the 
southeast edge of a unit rhombus that lies along the southeast edge of 
\HMreg{n}{2m}{R}{L}. Construct a similar path that begins at the midpoint of 
the southwest edge of $h_2$ and ends at the southwest edge of 
\HMreg{n}{2m}{R}{L}. 
\end{enumerate}
\begin{figure}[b]
\includegraphics[scale=0.6]{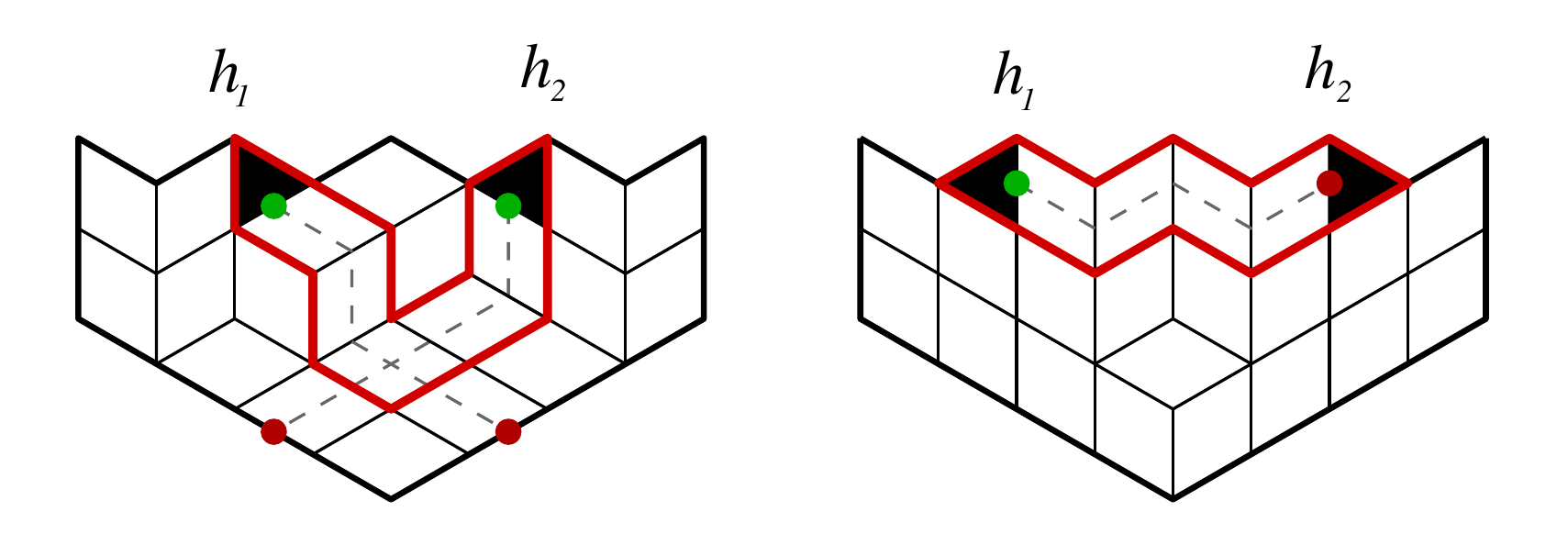}\caption{The propagation paths (outlined 
in red) between 
different configurations of holes $h_1$ and 
$h_2$.}\label{fig:propPaths}\end{figure}
It should be clear that one may always apply one of the above two steps. In 
the first case such a path certainly exists since in the translation of tilings 
to lattice paths, every tiling of $\HMreg{n}{2m}{R}{L}$ corresponds to a family 
of non-intersecting paths, one of which will begin at $h_1$ and end at $h_2$. 
In the second case, by translating a tiling to a set of paths across unit 
rhombi that begin at the southeast edge of \HMreg{n}{2m}{R}{L} one sees that 
there will always exist a path across rhombi from 
the southeast edge of \HMreg{n}{2m}{R}{L} that ends at $h_1$. Likewise, by 
translating the same tiling into paths across rhombi that instead begin at the 
southwest edge of \HMreg{n}{2m}{R}{L}, one sees that there will always exist a  
path across unit rhombi that ends at $h_2$. Note that these two paths will 
intersect at precisely one point, say $p_{1,2}$ (alternatively one could 
also say that the two paths across rhombi have in common precisely one unit 
rhombus).

The \emph{propagation path} from $h_1$ to $h_2$ is then the ribbon of unit 
rhombi that lie along the path from $h_1$ to $h_2$, either obtained 
directly from the path described by case (i), or in the second case by 
travelling from $h_1$ to $h_2$ via the unique intersection point $p_{1,2}$ (see 
Figure~\ref{fig:propPaths}).
\begin{figure}
\includegraphics[scale=0.6]{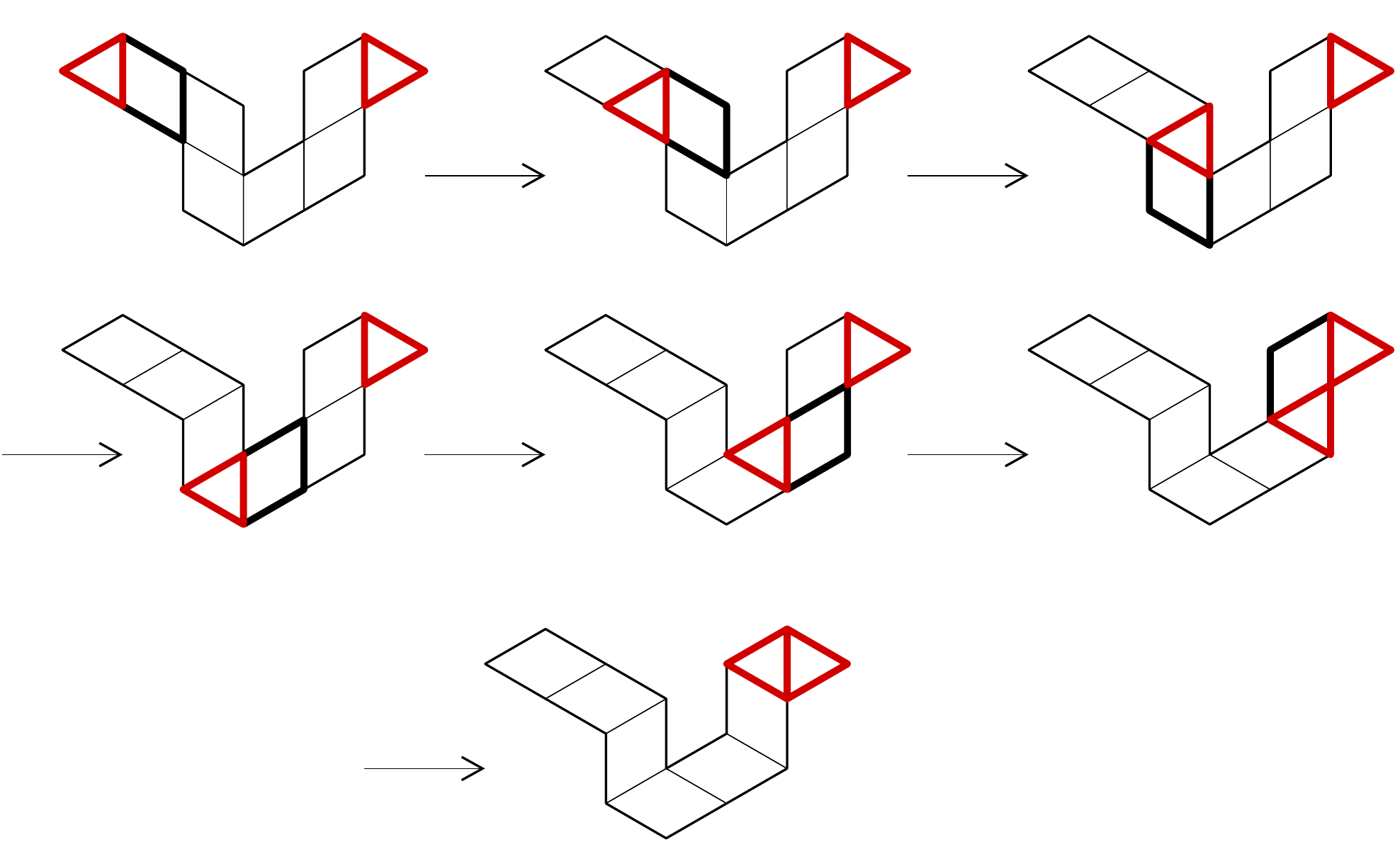}\caption{The transmission of a 
left pointing hole along a ribbon consisting of 6 unit 
rhombi, yielding a ribbon of 7 unit rhombi that contains 
no unit triangles.}\label{fig:swtiching2}
\end{figure}

The goal now is to transmit the unit triangle $h_1$ along the 
propagation path until it is in a position such that it shares an edge with 
the unit triangle $h_2$. One does this by successively interchanging the unit 
triangle $h_1$ with the neighbouring unit rhombus that lies 
between $h_1$ and $h_2$. 

This interchange is defined in the the following way: the 
unit rhombus that shares an edge with the unit triangle comprises a 
trapezium, so in order to interchange the two shapes one shifts the unit 
triangle one unit length along the longest edge of the trapezium. Such an 
interchange is demonstrated in the following figure:
\begin{figure}[H]
\includegraphics[scale=0.5]{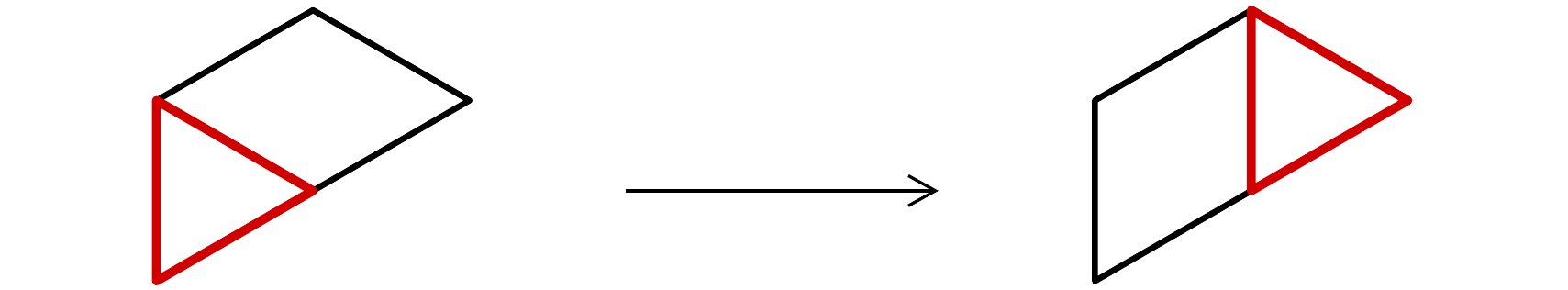}
\end{figure}\noindent(one may also see this as a reflection through a certain 
line of 
symmetry of 
the trapezium).

Given a propagation path that contains an arrangement of $k-1$ unit rhombi, 
where the ends of the ribbon are unit triangles, one may apply the interchange 
operation defined above 
$k-1$ times in order to transmit the leftmost hole $h_1$ along the path 
to the rightmost hole $h_2$. In doing so one transforms the original ribbon 
consisting of two triangular holes and $k-1$ rhombi into a ribbon of precisely 
the same shape, consisting instead of $k$ unit rhombi (the pair of 
triangular holes $h_1$ and $h_2$ that share an edge forms a hole that may 
be equivalently thought of as a unit rhombus). An example of the transmission 
of a hole along a ribbon of rhombi may be found in 
Figure~\ref{fig:swtiching2}. It should be clear that transmitting a triangular 
hole in this way only affects the rhombi that lie along the propagation path, 
thus any two distinct tilings of \HMreg{n}{2m}{R}{L} may be transformed under 
such an operation into two distinct tilings of 
$\HMreg{n}{2m}{\emptyset}{\emptyset}$.

The above argument establishes that the individual entries of the matrix 
$\widecheck{E}_{R,L}$ are bounded above by $1$, however one may easily extend 
such a mapping so that distinct tilings of \HMreg{n}{2m}{R}{L} containing a 
finite number of left and right pointing unit triangular holes may be 
transformed into distinct tilings of \HMreg{n}{2m}{\emptyset}{\emptyset}.
\begin{figure}\centering
\includegraphics[scale=0.5]{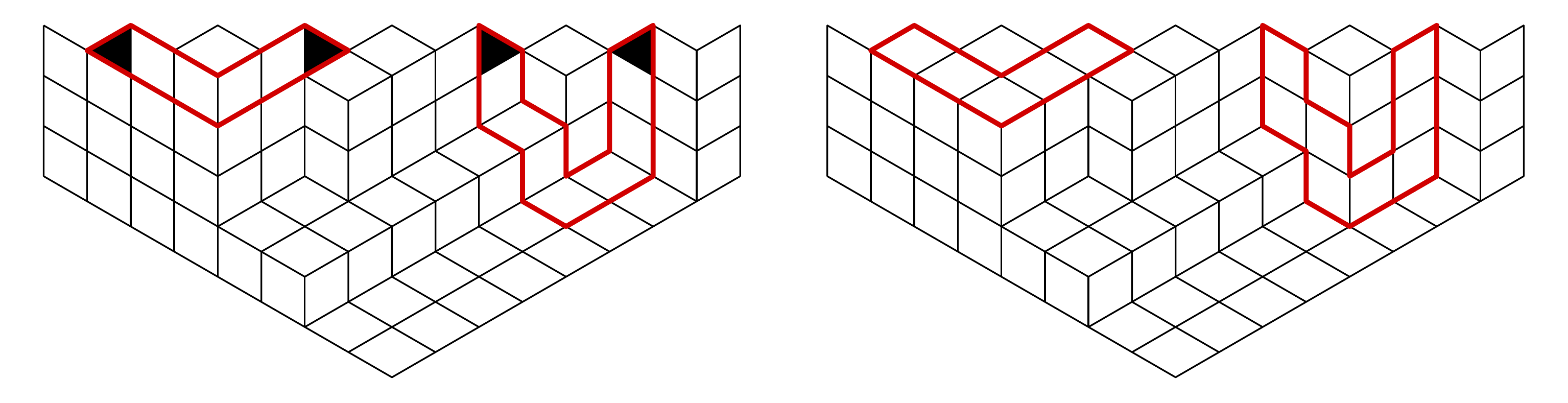}\caption{A tiling of 
\HMreg{8}{6}{\{-2,2\}}{\{-6,6\}} (left) and its 
corresponding unholey tiling of 
\HMreg{8}{6}{\emptyset}{\emptyset} under the map $\zeta$ (right). The sets 
$R=\{-2,2\}$ 
and $L=\{-6,6\}$ give rise to the tuple 
$(\textcolor{blue}{-6},-2,2,\textcolor{blue}{6})$, from 
which one obtains the set of ordered (and coloured) pairs of holes 
$\{(\textcolor{blue}{-6},-2),(2,
\textcolor{blue}{6})\}$ that determine the propagation paths along which one 
must apply $\zeta$.}\label{fig:Multip2}
\end{figure}

Given two sets $R$ and $L$ (where $|R|=|L|=k$) corresponding to multiple right 
and left pointing unit triangular holes along the zig-zag boundary of 
\HMreg{n}{2m}{R}{L}, colour all the elements of $R$ in one colour and all the 
elements of $L$ in another. Consider the tuple $T$ 
obtained by ordering all elements of $R\cup L$ in increasing order 
(irrespective of colour). From this tuple select the first pair 
of consecutive elements of differing colours that occur when $T$ is read from 
left to 
right, say $h_1$ and $h_2$. Remove these elements from $T$ and 
form a new tuple consisting of $2k-2$ 
coloured elements from $R\cup L\setminus\{h_1,h_{2}\}$, arranged in increasing 
order. Repeatedly applying 
this process ($k$ times in total) yields a set of 
ordered (and coloured) pairs $\{(h_1,h_2),\dots,(h_{2k-1},h_{2k})\}$ that 
determine the holes between which one should construct propagation paths. 
A moment's thought convinces oneself that selecting pairs of holes in 
this way ensures that no two distinct propagation paths between pairs 
of holes intersect, thus it is straightforward to see that under this extended 
mapping any two distinct tilings of \HMreg{n}{2m}{R}{L} give rise to two 
distinct tilings of \HMreg{n}{2m}{\emptyset}{\emptyset}. 
Figure~\ref{fig:Multip2} illustrates the unholey tiling obtained from a tiling 
of \HMreg{8}{6}{\{-2,2\}}{\{-6,6\}}.

What of a map between weighted tilings of 
\HPreg{n}{2m}{R}{L} and weighted tilings of 
\HPreg{n}{2m}{\emptyset}{\emptyset}? Observe that the holes that 
lie along the zig-zag boundary of \HPreg{n}{2m}{R}{L} are trapezia that lie 
within the peaks of the boundary, and such holes are equivalent to fixing a 
unit triangular hole within a ``fold" of the boundary, thus forcing a 
rhombus that shares 
with it a vertical edge. One may therefore apply precisely the same map $\zeta$ 
to tilings of \HPreg{n}{2m}{R}{L} to obtain tilings of 
\HPreg{n}{2m}{\emptyset}{\emptyset}. At first it would appear that one must be 
careful to 
consider the extra weighting of certain configurations, 
however after a little consideration it is straightforward to see that by 
transforming a weighted tiling of \HPreg{n}{2m}{R}{L} into a 
weighted tiling of \HPreg{n}{2m}{\emptyset}{\emptyset}, one only ever
increases the number of possible configurations of tiles that carry a combined 
weight of $2$. Thus the weight of a tiling of \HPreg{n}{2m}{R}{L} is at 
most the weight of its image under $\zeta$, hence the weighted count of all 
tilings of \HMreg{n}{2m}{R}{L} is at most the weighted count of all tilings of 
\HMreg{n}{2m}{\emptyset}{\emptyset}.
\end{proof}

\begin{cor}
The number of tilings of \Hreg{n}{2m}{R}{L} is at most the number of tilings of 
$H_{n,2m}$.
\end{cor}

\begin{rmk}\label{rmk:const1}
This corollary follows immediately from Theorem~\ref{thm:Inj}, however it 
is in fact a special case of a more general (conjectured) result, chiefly that 
the number of tilings of a hexagon that contains any amount of unit triangular 
holes distributed anywhere within its interior is at most the 
number of tilings of the same region without holes. An injective proof of this 
result will be the subject of future work by the author.
\end{rmk}

\section{Asymptotics}\label{sec:Asymp}
The goal of this final section is to establish asymptotic expressions for 
$\det \widehat{E}_{R,L}$ and $\det \widecheck{E}_{R,L}$ as the size of 
the boundaries of 
\HPreg{n}{2m}{R}{L} and \HMreg{n}{2m}{R}{L} tend to infinity and the distances 
between the holes grows large.

Suppose $\xi$ is a positive real number such that $2m\thicksim \xi n$. For 
a set of triangular holes determined by $R$ and $L$ within a sea of unit 
rhombi, the interaction between them is defined to be
$$\omega_H(R,L;\xi)=\lim_{n\rightarrow\infty}\frac{M(\Hreg{n}{\xi 
n}{R}{L})}{M(H_{n,\xi n})},$$
according to~\eqref{eq:CorrDef}. Inserting the result from 
Theorem~\ref{thm:ExactFull} into the right hand side 
above gives
$$\omega_H(R,L;\xi)=\lim_{n\to\infty}(\det \widecheck{E}_{R,L}\cdot\det 
\widehat{E}_{R,L}),$$
where $\widecheck{E}_{R,L}$ and $\widehat{E}_{R,L}$ are matrices of size $p$, 
independent 
of $n$. According to Theorem~\ref{thm:Inj} the product of these 
determinants 
is bounded above by $1$ for all $n$, thus one may consider the limit of 
the 
individual entries of each matrix as $n$ tends to infinity. 
To that end, consider a single entry of $\widecheck{E}_{R,L}$ for $r_j>l_i$,
\begin{multline*}\lim_{n\to\infty}\left(\pFq{4}{3}{\tfrac{r}{2}-\tfrac{n}{2}+1,1
,\tfrac
{r}{2}-\tfrac{l}{2}+2,\tfrac{n}{2}+\tfrac{r}{2}+\tfrac{1}{2}}{m+\tfrac{n}{2}+
\tfrac{r}{2}+2,\tfrac{r}{2}-m-\tfrac{n}{2}+2,\tfrac{r}{2}-\tfrac{l}{2}+\tfrac{3}
{2} }{1}\right.\\\times\frac { \Gamma 
(m+n+1) \Gamma 
(\frac{n}{2}+\frac{r_j}{2}+\frac{1}{2}) \Gamma (\frac{l_i}{2}+ m+\frac{n}{2}) 
\Gamma 
(m+\frac{n}{2}-\frac{r_j}{2}-1)}{\Gamma 
(\frac{n}{2}-\frac{r_j}{2}) \Gamma (m-\frac{l_i}{2}+\frac{n}{2}+1) 
\Gamma ( 
m+\frac{n}{2}+\frac{r_j}{2}+2))}\\\left.\times\frac{2^{r_j-l_i+2} \Gamma 
(m+\frac{3}{2}) 
\Gamma (\frac{n}{2} -\frac{l_i}{2}+\frac{1}{2}) }{\pi  (r_j-l_i+1) \Gamma (m) 
\Gamma 
(\frac{l_i}{2}+\frac{n}{2}) \Gamma (m+n-\frac{1}{2})}\right).\end{multline*}
Once again by Theorem~\ref{thm:Inj} the above expression is 
bounded above by $1$ for all $n$. Moreover since the sum is not alternating and 
terminates one may safely interchange the 
limit and sum operations as $n$ tends to infinity. Applying Stirling's 
approximation to the 
pre-factor above yields
$$\frac{\xi^{3/2} (\xi+2)^{3/2} 2^{r_j-l_i+2}}{\pi  
(r_j-l_i+1)(\xi+1)^{r_j-l_i+4}},$$
while the $_4F_3$ hypergeometric series reduces to
$$\pFq{2}{1}{1, 
\tfrac{r_j}{2}-\tfrac{l_i}{2}+2}{\tfrac{r_j}{2}-\tfrac{l_i}{2}+\tfrac{3}{2}}{
\frac { 1 } {
(\xi+1)^2}}.$$
As the distance between the holes grows (that is, the distance between the 
holes at $r_j$ and $l_i$ grows very large), 
this series reduces to a geometric series, hence
$$\check{e}_{i,j}\thicksim\frac{(\xi(\xi+2))^{1/2}}{
\pi(r_{j}-l_{i})}\left(\frac{2}{\xi+1}\right)^{r_j-l_i+2}$$ for holes that are 
far apart and point away from each other in a sea of unit rhombi.

Similarly if the holes point toward each other (that is, if $r_j<l_i$) then as 
$n$ tends to infinity one obtains
\begin{equation}\label{eq:echeckij}\check{e}_{i,j}\thicksim-\frac{2^{r_j-l_i+2}
(\xi(\xi+2))^ { 3/2 } } { 3\, \pi } \cdot\pFq { 2 } { 1 } { 2-\tfrac { l_i } { 
2 } +\tfrac { r_j}{2},\tfrac{3}{2}}{\tfrac{5}{2}}{-\xi(\xi+2)}.\end{equation}
Since the above hypergeometric series is terminating one may apply the 
following transformation formula 
$$\pFq{2}{1}{a,-n}{c}{z}=\frac{(1-z)^n(a)_n}{(c)_n}\pFq{2}{1}{-n,c-a}{1-a-n}{(1-
z)^{-1}}$$
(see~\cite[(1.8.10), with sum reversed on the right hand side]{Slat}) 
to~\eqref{eq:echeckij}, yielding
$$\check{e}_{i,j}\thicksim\frac{2^{r_j-l_i+2}(\xi(\xi+2))^{3/2}}{\pi(r_j-l_i+1)}
\pFq{2}{1}{1,\tfrac{r_j}{2}-\tfrac{l_i}{2}+2}{\tfrac{r_j}{2}-\tfrac{l_i}{2}
+\tfrac{3}{2}}{\frac{1}{(\xi+1)^2}}.$$
This agrees completely with the asymptotic approximation of $\check{e}_{i,j}$ 
for $r_j>l_i$, thus it follows that each entry of the matrix 
$\widecheck{E}_{R,L}$ is asymptotically
$$\check{e}_{i,j}\thicksim\frac{(\xi(\xi+2))^{1/2}}{
\pi(r_{j}-l_{i})}\left(\frac{2}{\xi+1}\right)^{r_j-l_i+2},$$
and a similar line of reasoning shows that each entry 
of 
$\widehat{E}_{R,L}$ is asymptotically
$$\hat{e}_{i,j}\thicksim\frac{(\xi(\xi+2))^{-1/2}}{
\pi(r_j-l_{i})}\left(\frac{2}{\xi+1}\right)^{r_j-l_i+2}.$$

For $\xi\neq1$ the interaction between holes either blows up or shrinks 
exponentially according to how they are positioned. More precisely, suppose at 
first that 
$\xi>1$. If the leftmost triangular hole lying within tilings of the plane is 
right pointing then as the distance between the holes becomes large the entries 
of an entire column of $\widehat{E}_{R,L}$ shrink exponentially, thus the 
determinant of $\widehat{E}_{R,L}$ also shrinks. If instead the leftmost hole 
is a left pointing triangle then the entries of an entire row blow up 
exponentially, thus the absolute value of the determinant grows exponentially 
large. It is easy to see that the same can be said of the matrix 
$\widecheck{E}_{R,L}$, and simlarly if $\xi<1$ then the converse is true.

Suppose now that $\xi=1$ and consider the $(i,j)$-entry of 
$\widecheck{E}_{R,L}$, that 
is,
$$\frac{1}{2\pi(x_i-y_j)},$$
where $x_i=-\frac{\sqrt{3}}{2}l_i$ and $y_j=-\frac{\sqrt{3}}{2}r_j$. It is 
clear that the determinant of $\widecheck{E}_{R,L}$ is the determinant of a 
certain Cauchy 
matrix, hence it has the following explicit evaluation
$$|\det 
\widecheck{E}_{R,L}|=\left(\frac{1}{2\pi}\right)^p\frac{\prod_{i=2}^p\prod_{j=1
}^{i-1}
d(r_i,r_j)d(l_i,l_j)}{\prod_{1\le i,j\le p}d(r_i,l_j)},$$
where $d(x,y)=\frac{\sqrt{3}}{2}|x-y|$ is the Euclidean distance between the 
midpoints of the vertical edges of the triangular holes at lattice distances 
$x$ 
and $y$ from the origin. 

Since the charge of each triangular hole is $\pm 2$, 
the right hand side of the above expression may be re-written as
\begin{equation}\label{eq:RUL}\left(\frac{1}{2\pi}\right)^p\prod_{1\le j<i\le 
|R\cup L|}d(h_i,h_j)^{\frac{1}{4}q(h_i)q(h_j)},\end{equation}
where $h_i,h_j\in R\cup L$. While this expression certainly holds for 
interleaving triangular holes of side length two it may be refined further in 
order to express the interaction as a product over holes of any even size.

Suppose $R$ and $L$ each contain a string of contiguous triangular holes 
$r_1,\dots,r_s$ and $l_1,\dots,l_t$ (respectively) that induce holes $h_1$ and 
$h_2$ of side lengths $2s$ and $2t$ (respectively) as described in 
Remark~\ref{Remark:Contig}. The charge of an induced hole is simply the 
sum of the charges of the holes of side length two that induce it 
(equivalently, the charge of each hole is $\pm$ twice the number of holes that 
induce it). As the distance between $h_1$ and $h_2$ grows large the distances 
between the individual triangular holes $r_1,\dots,r_s$ remain constant 
(similarly for $l_1,\dots,l_t$), thus
$$\frac{\prod_{i=2}^s\prod_{j=1}^{i-1}d(r_i,r_j)d(l_i,l_j)}{\prod_{i=1}^s 
\prod_{j=1}^td(r_i,l_j)}\thicksim\left(\prod_{h\in\{h_1,h_2\}}\prod_{i=0}^{\frac
{ 1 } { 2 } |q(h)|-1 } 3^ { i/2 } \Gamma(i+1)\right)\cdot 
d(h_1,h_2)^{\frac{1}{4}q(h_1)q(h_2)}.$$
Letting $\mathcal{H}$ index the set of triangular holes induced by $R$ and 
$L$ it follows from the above observation that~\eqref{eq:RUL} may be re-written 
as
\begin{equation}\label{eq:HPoverH}\prod_{h\in\mathcal{H}}\left(\prod_{s=0}^{
\frac { 1 } { 2 } |q(h)|-1 } \frac { 3^ { s/2 } \Gamma(s+1) } { \sqrt 
{2\pi}}\right)\prod_{1\le j<i\le 
|\mathcal{H}|}d(h_i,h_j)^{\frac{1}{4}q(h_i)q(h_j)},\end{equation}
for $h_i,h_j\in\mathcal{H}$, thereby establishing Theorem~\ref{thm:VertCor}.

Similar arguments may be used to show that
$$|\det 
\widecheck{E}_{R,L}|\thicksim\prod_{h\in\mathcal{H}}\left(\prod_{s=0}^{\frac{1}{
2}|q(h)|-1}
\frac{ 3^{(s+1)/2} \Gamma(s+1)}{\sqrt{2\pi}} \right)\prod_{1\le j<i\le 
|\mathcal{H}|}d(h_i,h_j)^{\frac{1}{4}q(h_i)q(h_j)}.$$
The product of the above expression together with~\eqref{eq:HPoverH} 
then yields Theorem~\ref{thm:MainThm}.
\pagebreak


\begin{thebibliography}{100}

\bibitem{PlanePart1}G.~E.~Andrews. \emph{Plane partitions I: The MacMahon
conjecture,} Studies in foundations and combinatorics, G.-C. Rota ed., Adv. in
Math. Suppl. Studies, Vol. 1 (1978), pp. 131--150.

\bibitem{CiucuFact}M.~Ciucu. \emph{Enumeration of perfect matchings in graphs
with reflective symmetry,} J. Combin. Theory Ser. A, Vol. 77 (1997), pp. 67--97.

\bibitem{CiuRot}M.~Ciucu. \emph{Rotational invariance of quadromer 
correlations on the hexagonal lattice,} Adv. in Math., Vol. 191 (2005), p. 46.

\bibitem{CiuRand}M.~Ciucu. \emph{A random tiling model for two dimensional 
electrostatics,} Memoirs of Amer. Math. Soc., Vol. 178 (2005), no. 839, 
pp. 1--106.

\bibitem{CiuScale}M.~Ciucu. \emph{The scaling limit of the correlation of holes 
on the triangular lattice with periodic boundary conditions,} Memoirs of Amer. 
Math. Soc. Vol. 199 (2009), no. 935, pp. 1--100.

\bibitem{CiuEmer}M.~Ciucu. \emph{ The emergence of the electrostatic field as a 
Feynman sum in random tilings with holes,} Trans. Amer. Math. Soc. Vol. 362 
(2010), pp. 4921--4954.

\bibitem{CiuPNAS}M.~Ciucu. \emph{Dimer packings with gaps and electrostatics,} 
Proc. Natl. Acad. Sci. USA, Vol. 105 (2008), pp. 2766--2772.

\bibitem{KrattFact}M.~Ciucu and C.~Krattenthaler. \emph{A factorization theorem
for
lozenge tilings of a hexagon with triangular holes,} arXiv pre-print, available
at {\tt http://arxiv.org/abs/1403.3323}.

\bibitem{CiuKrattInt}M.~Ciucu and C.~Krattenthaler. \emph{The interaction of a 
gap 
with a free boundary in a two dimensional dimer system,} Comm. Math. 
Phys. Vol. 302 (2011), pp. 253--289.

\bibitem{Feyn}R.~P.~Feynman. \emph{The Feynman Lectures on Physics, vol. II,} 
Addison-Wesley, Reading, Massachusetts, 1963.

\bibitem{FishSteph}M.~E.~Fisher and J.~Stephenson. \emph{Statistical mechanics 
of dimers on a plane lattice. II. Dimer correlations and monomers}, Phys. Rev. 
(2) Vol. 132 (1963), pp. 1411--1431.

\bibitem{LGV}I.~M.~Gessel and X.~G.~Viennot. {\it Determinants, paths, and plane
partitions}, preprint (1989), available at {\tt
http://people.brandeis.edu/~gessel/homepage/papers/pp.pdf}.

\bibitem{TG1}T.~Gilmore. \emph{Three interactions of holes in two dimensional
dimer systems}, arxiv preprint,~\email{arXiv:1501.05772}, available
at~{\tt http://arxiv.org/abs/1501.05772}.

\bibitem{GordPlane}B.~Gordon. \emph{Notes on plane partitions V,} J.
Comb. Theory Ser. B, Vol. 11 (2) (1971), pp. 157--168.

\bibitem{Keny1}R.~Kenyon. \emph{Local statistics of lattice dimers,} Ann. Inst. 
H. Poincar\'{e} Probab. Statist. Vol. 33 (1997), pp. 591--618.

\bibitem{KeOkSh}R.~Kenyon, A.~Okounkov, and S.~Sheffield. \emph{Dimers and 
amoebae,} Ann. of Math. Vol. 163 (2006), pp. 1019--1056.

\bibitem{Lind}B.~Lindstr\"{o}m. \emph{On the vector representation of reduced 
matroids}, Bull. London Math. Soc. Vol. 5 (1973), pp. 85--90.

\bibitem{CombAnal}P.~A.~MacMahon. \emph{Combinatory Analysis}, Vol. 2, Cambridge
University Press, 1916; (reprinted Chelsea, New York, 1960).

\bibitem{Proct}R.~A.~Proctor. \emph{Bruhat lattices, plane partitions
generating functions, and minuscule representations,} Europ. J. Combin. Vol. 5
(1984), pp. 331--350.

 \bibitem{ProppHoley}J.~Propp. \emph{Enumeration of matchings: Problems and 
progress,} New Persp. in Alg. Comb., L.~Billera,
A.~Bj\"{o}rner, C.~Greene, R.~Simion, and R.~P.~Stanley, eds., Mathematical 
Sciences Research Institute Publications, 38, Cambridge University Press 
(1999), pp. 255--291.

\bibitem{Slat}J.~L.~Slater. \emph{Generalized hypergeometric functions,}
Cambridge University Press (1966).

\bibitem{Wiki}Wikipedia. \emph{Method of images,} available at~{\tt 
https://en.wikipedia.org/wiki/Method\_of\_images}.
\end{thebibliography}
 \end{document}